# The properties of resistive MHD modes and unstable spectra in advanced tokamak regimes

M. Coste-Sarguet[1,*] and J. P. Graves[1,2]

[1] *École Polytechnique Fédérale de Lausanne (EPFL), Swiss Plasma Center (SPC), CH-1015 Lausanne, Switzerland*

[2] *York Plasma Institute, Department of Physics, University of York, York, Heslington, YO10 5DD, United Kingdom*

Advanced tokamak regimes, featuring extended regions of low magnetic shear, are promising candidates for future fusion reactors but are also more prone to specific kinds of MHD instabilities. The proximity to a rational surface in a very low shear region weakens field line bending stabilisation and amplifies the effects of toroidal coupling between modes, leading to the emergence of long-wavelength resistive infernal modes. These modes can grow collectively as a discrete spectrum, leading to a cascade of different perturbations for single mode numbers $(m, n)$, with subdominant modes showing increasingly oscillatory radial structures. These spectra of fast-growing modes are significant for developing stable scenarios in future reactors, and for the understanding of global reconnection events like sawteeth, motivating a deeper investigation into their fundamental physics. Deriving new analytic solutions, including a generalisation of the ideal interchange dispersion relation to non monotonic $q$ profiles, and extending a modular linear resistive MHD solver, we investigate how resistivity, compressibility, toroidal effects, and shaping influence stability, especially in reversed shear $q$ profiles. It is also shown that common assumptions in numerical calculations prevent the observation of the full variety of modes present in these advanced scenarios.

## I. INTRODUCTION

A new class of hybrid or advanced scenarios with improved confinement properties and low magnetic shear in the core (see Fig. 1) is a promising prospect for ITER and DEMO. Indeed the last record-breaking DT pulses at JET were achieved in hybrid mode. However, having an extended region of low magnetic shear near a rational surface greatly weakens the field line bending stabilisation and enhances the effect of toroidal coupling between modes. This coupling give rise to broad perturbations called infernal modes, which can extend over all the low shear zone.

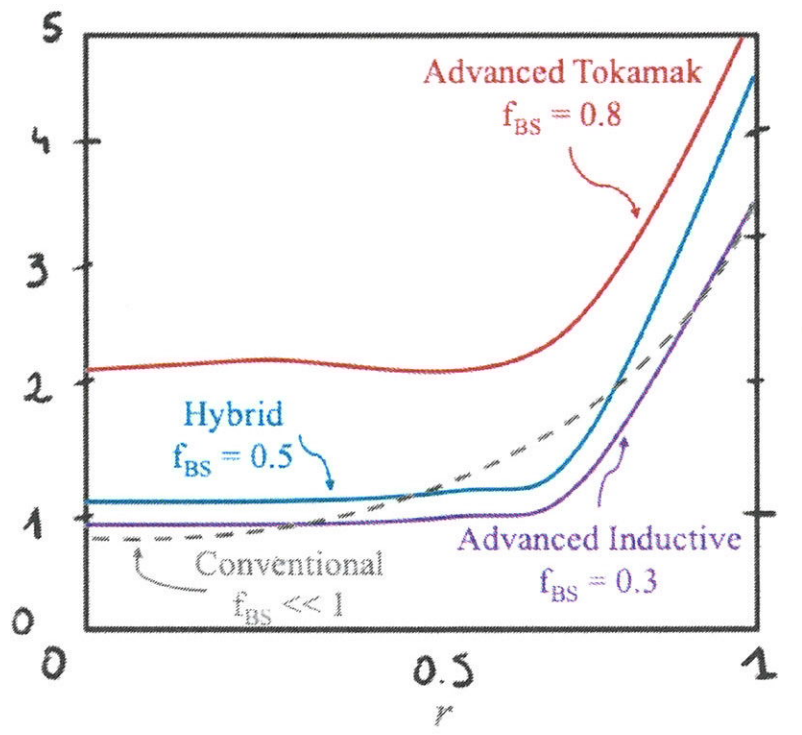

Figure 1: Different kinds of advanced tokamak scenarios, and the typical associated bootstrap current fraction $f_{BS}$.

* Corresponding author: margot.coste-sarguet@epfl.ch

Studying infernal modes requires keeping higher order toroidal corrections often overlooked for more sheared, conventional scenarios. In Reference [1], a unified and global description of linear toroidal resistive MHD instabilities including these infernal corrections was derived. The results were obtained in a resistive MHD framework, assuming a static equilibrium, circular flux surfaces and within straight field line coordinates and an inverse aspect ratio expansion.

In Reference [2], we presented and benckmarked a modular eigenvalue solver implementing these equations. We then used it to explore fundamental physics linked with very low shear $q$ profiles. An outcome of this study was the identification of an unstable spectrum of resistive and ideal infernal modes. Meaning that for single mode numbers $m$ and $n$ a cascade of different perturbations can occur. Subdominant modes with lower growth rates oscillate with shorter radial wave-length, and can be characterised by Bessel functions for the ideal case (the analytic description of such a spectrum for a $q \neq 1$ rational surface was made in [3]). The modular solver was required for cases where the $q$ profile crossed the rational surface, exhibiting that a wide cascade of unstable resistive infernal modes are produced at low beta.

The current paper aims to make use of this unified stability description and modular solver, exploring the parameters affecting the ideal and resistive stability of advanced scenarios, especially reverse shear ones. Indeed, most of the current literature on such scenarios focuses on double tearing modes, while there is limited research on the linear core stability to resistive infernal modes. References [4] (TPX scenarios) and [5] provide $\beta$ limits for different equilibria, but they do not investigate the influence of shaping or closely examine the type of modes that arise, simply suggesting that that resistive infernal modes may be linked to double tearing modes. A key fo-cus of the current work is to examine the mode structure and how *e.g.* the tearing mode branch can transform into the internal kink or resistive infernal branch. For non-resonant scenarios on the other hand (ideal physics), we aim to obtain analytic criteria and dispersion relations, using the modular solver as both guidance and validation. Reference [6] explored the impact of shaping and studied the linear and nonlinear evolution of internal modes for low and reverse-shear $q$ profiles, around a $q = 1$ surface. Our work builds on this by examining the spectrum of unstable modes associated with resistive effects and shaping. More recently [7] used global gyrokinetic simulations to study the kinetic infernal mode in reverse shear scenarios. Our study remains within the MHD framework, including resistive diffusion, compressibility, and shaping effects in toroidal geometry to further assess their impact on stability.

In the first section, we begin by presenting how basic shaping effects are included in the model. We then show ideal results from the full equations, as well as simplified analytic expressions, including a generalisation of the interchange dispersion relation to reverse shear case, to predict the existence of a core instability spectrum, studying the effects of shaping and magnetic shear. In the second section we explore how resistivity, negative triangularity, and infernal effects interact. Finally, we examine how common simplifying assumptions in many numerical codes can limit the observation of the full range of modes arising in these scenarios.

## II. ANALYTIC PREDICTIONS FOR IDEAL MHD WITH SHAPING

We generalize the model and modular solver introduced in [1] and [2] by relaxing the circular flux surfaces assumption, introducing leading order shaping effects.

***Toroidal corrections: modification of the Mercier term*** We consider flux surfaces with constant elongation $\kappa(r) = \kappa_1$ along the minor radius $r$, and triangularity following $\delta(r) = \delta_1\, r/r_1$ for $r < r_1$, where $r_1$ marks the end of the low shear region. Such radial dependence is exact in a large aspect ratio, low beta tokamak, for a locally shearless $q$ profile and generally provides a good approximation otherwise (see Appendix A for more details). The change in the ideal Mercier criterion $D_I$ ($D_I > 1/4$ corresponding to an ideal interchange unstable configuration) due to shaping in the presence of weak magnetic shear is [8]:

$$s^2 D_I(r) = \alpha\epsilon\left[\frac{1}{q_s^2} - 1 + \frac{3}{4}(\kappa-1)\left(1-2\frac{\delta}{\epsilon}\right)\right], \quad (1)$$

In the present work, $s = rq'/q$ is the magnetic shear, $\alpha(r) = -\frac{2\mu_0 q^2 R_0}{B_0^2}\frac{\mathrm{d}P}{\mathrm{d}r}$ characterises the pressure drive, and $\epsilon = r/R_0$ is the inverse aspect ratio. The safety factor at the rational surface is denoted $q_s = q(r_s) = m/n$, where $m$ and $n$ are the poloidal and toroidal mode numbers. We define the elongation $\kappa$ and the triangularity $\delta$ as in [8]. Our model equations include explicit coupling to the two nearest poloidal sidebands. Higher order sidebands $m\pm2$, $m\pm3$ are included in a large aspect ratio expansion with shaping effects, and these terms are implicitly embedded in (1) as seen *e.g.* in Ref. [9]. With $\kappa > 1$ and $\delta < 0$, both ideal and resistive interchange instabilities can now occur even in cases where $q_s = 1$ and even $q_s > 1$, due to the effect of shaping on average curvature.

***Neglect quasicylindrical corrections*** Additional shaping contributions [10], so-called quasicylindrical effects, are not included in the change of Mercier term. Indeed, the analysis of [10] was restricted to $m = 1$ cases. Given the length of the calculations needed to generalise these terms to arbitrary $m$, we restrict ourselves to a parameter space where these corrections can be safely neglected. The quasicylindrical corrections are terms in $O(e^2\Delta q^3)$, where $e = (\kappa-1)(\kappa+1)$ (see in Refs. [11] and [12]) and $\Delta q = q - q_s$ in the core region. These terms are analogous to the toroidal corrections in $\Delta q$ of Bussac, which we derived and included for circular cross sections in eq. (23) of Ref. [2]. The quasicylindrical shaping terms become important when the Mercier term vanishes, as in the $q_s = 1$ case discussed in that previous work especially if $\Delta q$ is large. In that context, these toroidal $\Delta q$ corrections allowed to observe the full spectrum of instabilities, even when $q_s = 1$. However, if the flux surfaces are shaped, the situation changes: the Mercier term (1) no longer vanishes, even for $q_s = 1$. In this case, the toroidal shaping corrections will always dominate, especially for the cases with $\Delta q \sim \epsilon$, or even much smaller. The quasicylindrical corrections may compete only for $\Delta q \sim 1$, the case for which there were originally derived. Additionally, Ref. [12] closely examined the implications of including or not these quasicylindrical corrections for an internal kink mode. All the terms are safely ignored in the current manuscript, even the quasicylindrical triangularity term in $O(\delta^2\Delta q)$ for cases where $\Delta q << 1$, and with small to moderate $\delta/\epsilon$.

***Change in governing equation & modular solver*** The full set of equations solved by the modular solver is provided in the Appendix B, including the $s^2 D_I$ contribution in (1). The complete expression for the resistive Mercier term is given in eq. (5.24) of [13]. Considering weak magnetic shear (which is consistent with our assumption on the $r$ dependence of $\kappa(r)$ and $\delta(r)$) and moderate shaping, we neglect terms of order $O(rq'e)$. Under these approximations, shaping does not contribute additional resistive terms. Neglecting the quasicylindrical corrections, the governing equation for the main harmonic of the radial plasma displacement is thus modified according to the change in the ideal Mercier term, as given in (1). This corresponds to equation (B2) for $m \neq 1$, or equation (B3) for $m = 1$.

***Obtaining a dispersion relation for non resonant scenarios*** We stress that advanced tokamaks scenarios often display a wide region of low magnetic shear in the core. Several attempts have been made to derive analytic exact solutions to the ideal MHD stability equation for such equilibrium. For the $q_s = 1$ case, this was done by Waelbrock and Hazeltine [14], or Hastie and Hender [15]. Solutions to the $q_s \neq 1$ cases were found by Wahlberg and Graves in [3], uncovering spectra solutions for a single pair of mode numbers. The latter work considered the low shear region as having a flat $q$ profile. In [2], we kept the same simplified form for the $q$ profile, and we retained the higher order toroidal corrections (toroidal $\Delta q$ terms mentioned in the previous section), absent in all of the works cited above. A dispersion relation valid for non-resonant $q$ profiles was obtained, as well as an explicit expression for the highest growth rate. Keeping these 'infernal' corrections not only allowed us to improve the accuracy of the marginal stability conditions, but also to obtain general results showing a spectrum of solutions arising for $q_s \neq 1$, but also for $q_s = 1$.

In this section, we further generalise the dispersion relation obtained in [2]. First we add shaping, and look at the evolution of the spectrum of solutions with elongation and triangularity. Then we add magnetic shear, extending the validity of the description to resonant (monotonic and reverse shear) $q$ profiles. The assumption of a parabolic pressure profile is maintained throughout.

### A. Dispersion relation & stability criteria for monotonic non-resonant $q$ profiles with shaping

A first generalisation is to relax the circular flux surface assumption, by introducing basic shaping effects. We define the dimensionless, $r$-independent analogue of $D_I(r)$:

$$\widehat{D_I} = s^2 D_I(r)\frac{r_\star^2}{r^2} = \left(\frac{\alpha}{r}\right)\frac{r_\star^2}{R_0}\left[\frac{1}{q_s^2} - 1 + \frac{3}{4}(\kappa-1)\left(1-2\frac{\delta}{\epsilon}\right)\right], \tag{2}$$

where $r^*$ is a general notation, indicating either the end of the low shear region for monotonic profiles (so in this section $r^* = r_1$), or the reversal point for reversed $q$ profiles. The expression (2) will be useful throughout this section to quantify Mercier effects in configurations where the magnetic shear vanishes globally or locally.

#### 1. *The generalised dispersion relation & shaping effects on the spectrum*

We make use of the modified Mercier term in eq. (1), to modify the dispersion relation that we obtained in [2]:

$$D(\gamma^2/\omega_A^2) = \frac{\sigma r_1^{2m+2}}{\lambda}\left[\frac{J_{m+1}(x)}{xJ_m(x)} - \frac{1}{2m+2}\right] - 1 = 0\,,$$
$$\text{with } x = \sqrt{\frac{\lambda r_1^2}{Q}}, \tag{3}$$

and with $Q = \frac{\gamma^2(1+2q_s^2)}{m^2\omega_A^2} + \frac{\Delta q^2}{q_s^4}$, $\sigma = \frac{\alpha}{2rq_s^2}C_0^+$ (same definition as in Ref. [3]), $r_1$ the end of low shear region, as well as:

$$\lambda = -\frac{\alpha}{rq_s^2R_0}\left[1 - \frac{1}{q_s^2} - \frac{3}{4}(\kappa-1)\left(1-2\frac{\delta}{\epsilon}\right)\right] + \frac{\Delta q}{q_s^3}\left[\frac{1}{R_0^2}\left(\frac{13}{4}\right) + \frac{3}{4}\frac{\alpha^2}{r^2}\right]$$
$$= +\frac{\widehat{D_I}}{q_s^2 r_\star^2} + \frac{\Delta q}{q_s^3}\left[\frac{1}{R_0^2}\left(\frac{13}{4}\right) + \frac{3}{4}\frac{\alpha^2}{r^2}\right], \tag{4}$$

note that the $3/4\,\alpha^2/r^2$ term vanishes if there is exact resonance (details will be provided in a future publication). Since $\Delta q$ is assumed to be constant in this derivation, and with $\alpha/r$ also constant for a parabolic pressure profile, $x$ is constant as well and the dispersion relation (3) does not depend on $r$. This dispersion relation involves Bessel functions, typical of the higher oscillating solutions. It reveals a solution for the linear growth rate $\gamma^2/\omega_A^2$ (normalised to the Alfvèn frequency $\omega_A$), through $x$ and $Q$, as well as the eigenfunctions (radial plasma displacement):

$$\xi = \frac{\sigma}{\lambda}\left(\frac{r_1^m J_m[k(\gamma/\omega_A)r]}{rJ_m[k(\gamma/\omega_A)r_1]} - r^{m-1}\right),$$
$$\text{with } \quad k(\gamma/\omega_A) = \sqrt{\lambda/Q} = \sqrt{\lambda\left(\frac{\gamma^2(1+2q_s^2)}{m^2\omega_A^2} + \frac{\Delta q^2}{q_s^4}\right)^{-1}}. \tag{5}$$

Figure 2 plots the dispersion relation (3), for a $m/n = 2/2$ internal kink case. A typical hybrid $q$ profile is chosen, just above the rational with $q_0 = 1 + 1\cdot 10^{-4}$ ($q_0 = q(r=0)$ being the value of $q$ at the magnetic axis). The $q$ profile is flat in the core until $r_1 = 0.4$, and then grows quadratically until $q_a = 4$. We recall that on this plot, every root of the dispersion relation corresponds to a linear growth rate. Several roots thus means a spectrum of instabilities for single mode numbers $m$ and $n$. We observe that shaping can drastically change the number of modes. In the top panel we can see how operationally favourable shaping (here positive triangularity with $\kappa = 1.7$ and $\delta_1 = +0.08$) makes the spectrum from the unshaped equilibrium vanish. Only the most unstable mode remains. In contrast, the bottom panel shows that with unfavourable shaping (here elongation only with $\kappa = 1.7$), a broad spectrum of unstable modes reappears. Figure 3 shows the plasma displacements for the same elongated equilibrium ($\kappa = 1.7$) as in the right panel of Fig. 2. We restrict ourselves to the 6 most unstable modes, yet we still observe how the Bessel functions solution for the displacement (eq. (5), plotted on the left in Fig. 3) provides a very good approximation to the plasma displacements obtained by the modular solver (*i.e.* solving the full equations, shown on the right). As expected, the solver's displacements are mostly confined to the low-shear zone, within $r = 0.4$, particularly for the lower growth rate modes. There is also good agreement between the predicted growth rates, obtained from the roots of the dispersion relation (left), and those from the solver (right). This consistency reinforces our confidence in using these analytical results in the rest of this section.

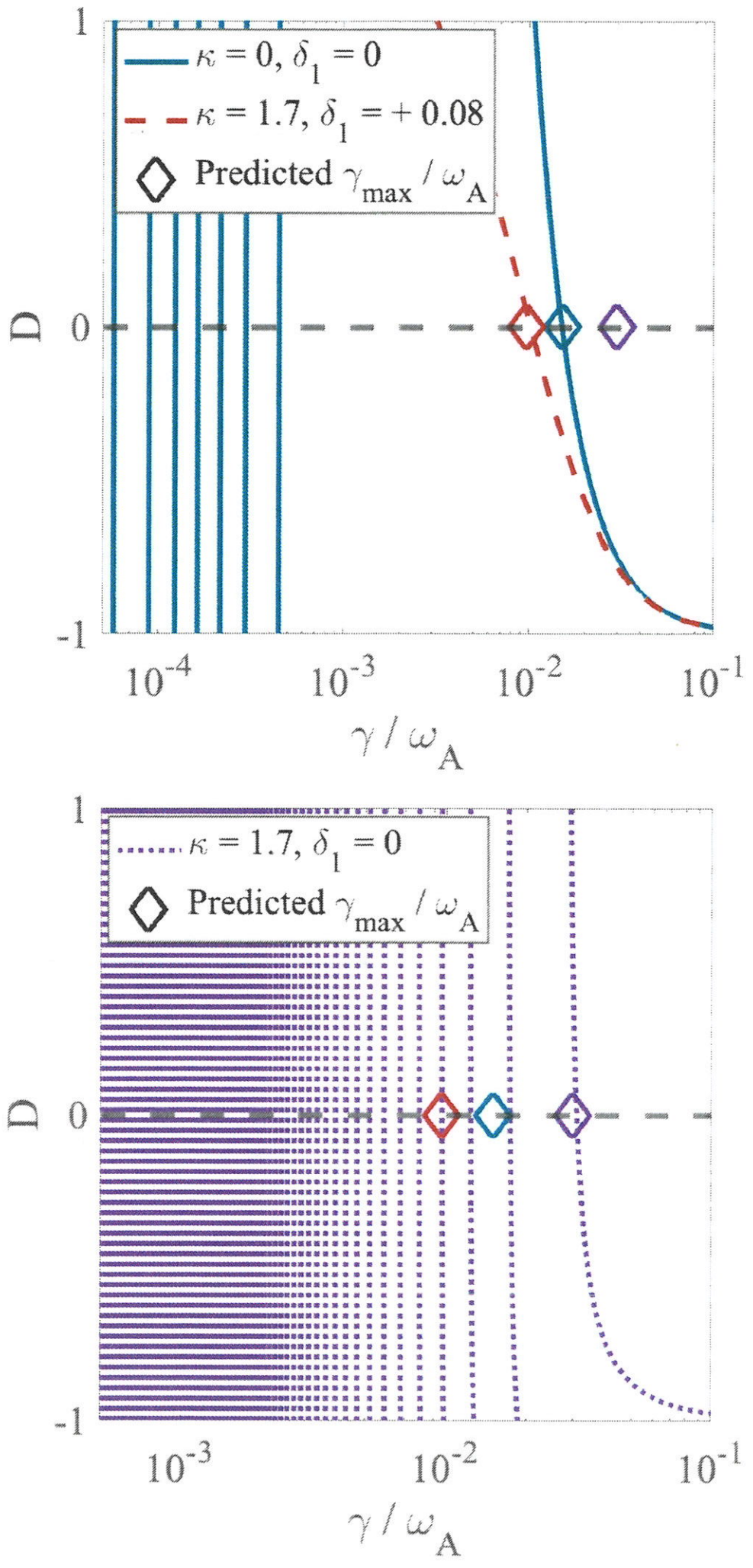

Figure 2: Dispersion relation (3) $D(\gamma^2/\omega_A)$, for a $m/n = 2/2$ mode with or without shaping. Hybrid $q$ ($\Delta q = 1\cdot 10^{-4}$) and parabolic pressure profile with $\alpha(0.4) = 0.2$ (equivalently $\beta_p(0.4) = 0.4$). Diamond markers indicate the analytically predicted growth rates from Eq. (7).

### 2. *Growth rate of the most unstable mode*

To obtain simplified expressions and instability criteria, we express $\sigma$ in terms of the convenient quantity $\Lambda_{m,n}$, which is linked to the sideband perturbation. Following Ref. [16], we can write:

$$\Lambda_{m,n} = \frac{(m+1)(m+2)}{2m(2m+1)}\left(\frac{q_a}{q_s} - 1\right)\left(\frac{r_1}{1-r_1}\right). \tag{6}$$

This expression was derived under the assumption that $r_1 \ll 1$, and for a $q$ profile that is constant for $r < r_1$, then grows quadratically up to $r = a = 1$. Since it remains a good analytical approximation for other low-shear profiles as well, we will use it throughout this section.

We can derive the growth rate of the most unstable mode, generalizing the expression from Ref. [2] (eq. (26)) to include shaping effects:

$$\frac{\gamma^2}{\omega_A^2} = \frac{m^2}{1+2q_s^2}\left[\frac{(\alpha/r)^2 r_1^2 \Lambda_{m,n}}{8(1+m)^2(2+m)} + \frac{1}{2(1+m)(3+m)}\left(\frac{\widehat{D_I}}{q_s^2} + r_1^2\frac{\Delta q}{q_s^3}\left[\frac{13}{4}\frac{1}{R_0^2} + \frac{3}{4}\frac{\alpha^2}{r^2}\right]\right) - \frac{\Delta q^2}{q_s^4}\right]. \tag{7}$$

It is plotted in Fig 2, and we observe excellent agreement between the explicit analytical prediction and the dispersion relation root. There is also excellent agreement with the exact solution obtained with the modular solver, as seen by the growth rates shown in Figure 3.

### 3. *Modified Mercier criterion for interchange modes in zero shear cases*

In configurations with either $q_s \neq 1$ or non-circular flux surfaces (interchange unstable cases), the Mercier term is nonzero and the $\Delta q$ corrections in $\lambda$ (Eq. (4)) can be

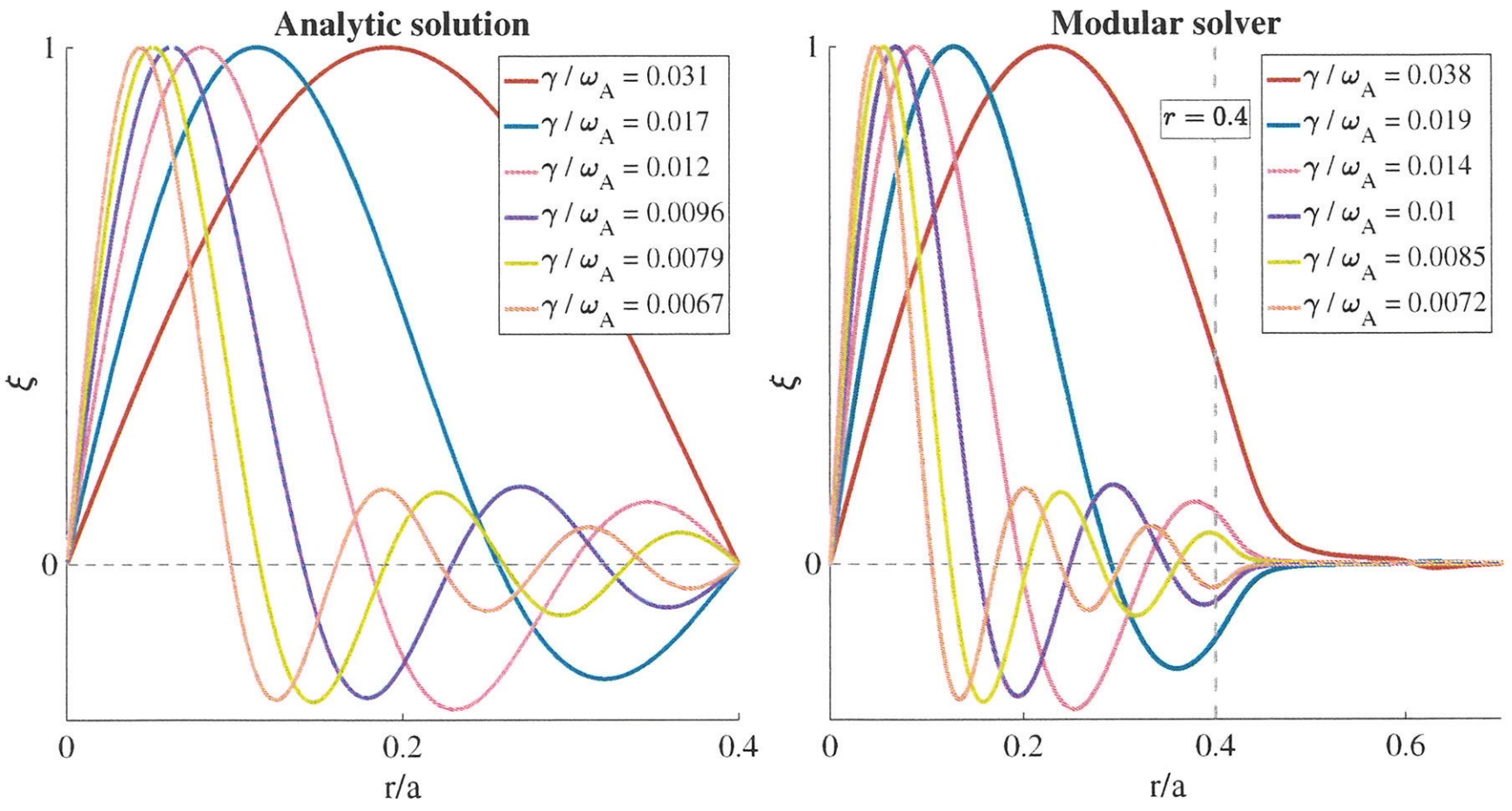

Figure 3: Radial plasma displacements for the 6 most unstable $m/n = 2/2$ modes, and their associated growth rates. The analytical solution is plotted on the left (eigenfunctions given by eq. (5)), and the modular solver results are shown on the right. Same equilibrium as for Fig. 2 the right: $q = q_0 = 1 + 1 \cdot 10^{-4}$ constant until $r = r_1 = 0.4$, $\kappa = 1.7$, $\delta_1 = 0$ and $\alpha(r = 0.4) = 0.2$.

neglected. The simplified result is:

$$\lambda_0 = -\frac{\alpha}{r q_s^2 R_0}\left[1 - \frac{1}{q_s^2} - \frac{3}{4}(\kappa - 1)\left(1 - 2\frac{\delta}{\epsilon}\right)\right] = +\frac{\widehat{D_I}}{q_s^2 r_\star^2}. \quad (8)$$

The growth rate of the most unstable mode is now given by:

$$\frac{\gamma^2}{\omega_A^2} = \frac{m^2}{1 + 2q_s^2}\left[\frac{\widehat{D_I}}{2q_s^2(1+m)(3+m)} - \frac{\Delta q^2}{q_s^4}\right], \quad (9)$$

where $\Lambda_{m,n}$ no longer appears, as the infernal coupling to sidebands is neglected here. The associated marginal stability condition is given below.

Interchange instability criterion in the no shear limit:

$$\frac{\Delta q^2}{q_s^2} 2(1+m)(3+m) < \widehat{D_I}\,, \quad (10)$$

where the left-hand side is the stabilising field line bending contribution, while the right-hand side contains the pressure drive $\widehat{D_I}$, which will be either stabilising ($\widehat{D_I} < 0$) or destabilising ($\widehat{D_I} > 0$). This modified Mercier criterion offers a compact way to assess when the most unstable mode is expected to grow in the absence of magnetic shear.

## B. Dispersion relation with shear and shaping

The dispersion relation (3) and the modified Mercier criterion (10) were derived assuming a constant $q$ profile in the core. As such, they are not well suited for studying for instance reversed shear cases, which can also display interesting instability spectra under certain conditions (see for instance Figure 4a for a $m = n = 1$ spectrum in a reverse shear elongated equilibrium). We would like to also be able to describe such spectra and derive an instability criterion for more general equilibria. When retaining core shear effects in the original governing equation for the growth rate and displacement (eq. (17) in Ref. [2]), it is no longer possible to obtain closed solutions to this generalised problem. An idea to get around this problem is to obtain the growth rates by minimization of the energy and asymptotic matching to an analytic expansion close to the minimum of $q$. We continue to focus on cases dominated by interchange terms, due to non-zero average curvature, modified by shaping, with small but non-zero infernal contribution.

### 1. $\delta W$ approach

We will break the normalised energy into two different contributions (following Ref [1], eq. 115):

$$\frac{1}{\xi_0 r_\star^{m+1}}\left(r^{2m+1}F(r)^2\frac{\mathrm{d}}{\mathrm{d}r}\left(r^{1-m}\xi^{(m)}\right)\right)\Bigg|_{r=0}^{r=r^*} = \delta W\,, \quad (11)$$

where $\xi_0 = \lim_{\delta_r \to 0^+} \xi^{(m)}(r_\star - \delta_r)$, and $F(r)^2 = \frac{\gamma^2(1+2q_s^2)}{m^2\omega_A^2} + \left(\frac{1}{q(r)} - \frac{1}{q_s}\right)^2$. The $\star$ subscript indicates that we take the value at $r_\star = r_s$ for a monotonic $q$ profile, and at $r_\star = r_{min}$, where $q(r_{min}) = q_{min}$ for a reversed $q$ profile. This is chosen in order to have a suitable expansion around $r_\star$ of the $q$ profile. The left hand side of (11) corresponds to inertia and leading order field line bending terms, and the right hand side is given by:

$$\delta W = \frac{1}{\xi_0}\int_0^{r_\star} dr\left[(1-m)\frac{r^{m+1}}{r_\star^{m+1}}\xi^{(m)}\frac{\mathrm{d}}{\mathrm{d}r}\left(\left(\frac{1}{q(r)}-\frac{1}{q_s}\right)^2\right)-\frac{r^{m+2}}{r_\star^{m+3}q_s^2}\widehat{D_I}\xi^{(m)}\right], \tag{12}$$

where the first term in $\delta W$ is associated with magnetic field line bending with magnetic shear. We move it to the right side of equation (11) to obtain a convenient closed integral on the left hand side, as in eq. (116) in Ref. [16]. Indeed this term depends on shear, which is weak, and does not involve derivatives of $\xi$, which are expected to be the quantities that are the most relevant in the layer. It is thus justified to include this term in $\delta W$, and adopt a global ansatz solution for $\xi$ to evaluate it. This expression (12) for $\delta W$ corresponds to the original equation for $\xi$ (equation (17) in [2]), now with the inclusion of shaping effects in the Mercier term $\widehat{D_I}$ (1), as described in the previous subsection. In this expression, we retain only the interchange terms since we investigate shaped cases where interchange terms dominate over infernal terms and other toroidal terms proportional to $\Delta q$. This corresponds to what we just did in section II A 3, for constant $q$ profiles. Now we generalize to non-zero shear cases. This simplification is also justified by observations in Ref. [2], where it was found that, for the higher oscillating modes in the spectrum, higher order infernal corrections have a negligible impact on the growth rate, with the interchange contributions dominating. This approximation holds even more strongly in the case of shaped equilibria, and additionally leads to a more manageable final expression.

For the left-hand side of eq. (11), we do not use an ansatz for $\xi$, as this term is sensitive to what happens in the layer where inertia is important. We define the dimensionless layer variable $x \equiv \frac{r-r^*}{r^*}$, and we express $q$ as: $q = q_\star(1+s_\star x+(q_\star'' r_\star^2/2q_\star)x^2)$ which allows for monotonic and non-monotonic $q$ profiles. The details of the calculation can be found in Appendix C, and we obtain:

$$-\frac{\sqrt{\hat{\gamma}^2+\frac{\Delta q_\star^2}{q_\star^2 q_s^2}}\sqrt{\frac{s_\star^2}{q_\star^2}+\frac{\Delta q_\star q_\star'' r_\star^2}{q_s^2 q_\star^2}-\frac{2\Delta q_\star s_\star^2}{q_s^2 q_\star}}}{\pi} = \delta W\,, \tag{13}$$

with $\hat{\gamma} = \frac{\gamma\sqrt{1+2q_s^2}}{m\omega_A}$. $\Delta q_\star \equiv q(r_\star) - q_s$, so for a monotonic resonant profile we have $\Delta q_\star = 0$.

***Checking expression*** (13) The expression (13) can be checked against several known results. For the monotonic case, $r_\star = r_s$, so $\Delta q_\star = 0$, and $s^* \neq 0$, and equation (13) becomes:

$$-\frac{\hat{\gamma}}{\pi}\frac{s(r_s)}{q_s} = \delta W\,,$$

which is the same as the $m = 1$ version in equation (101) from [1]. Reference [17] studied the energy minimization problem for a variety of $q$ profiles with $q_s = 1$ surfaces, for $m = 1$ modes. We can compare our result with their equation (14), which applies to non-monotonic $q$ profiles where $q_{min} \neq 1$. In our derivation, we retained less terms in the singular layer calculation, stopping at $O(x^2)$ when expanding $\left(\frac{1}{q}-\frac{1}{q_s}\right)$. To recover our approximation, we consider only terms up to $O(x^2)$ from the result in [17], which becomes:

$$\bar{\gamma} = \frac{1}{(r_1^2 q''\delta q)^{1/2}}\left(-\pi\frac{r_1^2}{R^2}\delta W^T\right),$$

with $\bar{\gamma} = \sqrt{3\gamma^2/\omega_A^2+\Delta q^2}$. This expression is the same as our result (13) (with $s_\star = 0$), with the different normalisation convention for $\delta W$ ($\delta W = \frac{r_1^2}{R^2}\delta W^T$).

### 2. *Ansatz choice to compute $\delta W$*

From the singular layer calculation, we have the expression linking the potential energy $\delta W$ to the growth rate : equation (13). However, we still need an explicit expression for $\delta W$. The usual strategy to compute the contribution from this potential energy (12) is to use an ansatz/global solution for $\xi$, since it does not depend on the specific details of the inertia layer. A common and sensible ansatz is the modified Heaviside step function: $\xi(r) = \xi_0\left(\frac{r}{r_\star}\right)^{m-1}H(r_\star - r)$. Using it would discard all the information about a potential spectrum of instabilities, which is precisely what we want to describe here. Another approach, which we adopt in this work, is to start from a different ansatz. We make use of the fact that we want to describe modes occurring in nearly flat $q$ profile configurations. Thanks to the constant $q$ calculations outlined in the previous subsection, we already have a good understanding of how the short wavelength oscillations modes will behave in such equilibria. Therefore, we will use the plasma displacement expression provided in (5). This expression is a valid solution of the shearless ideal ODE for $\xi$ (eq. 22 in Ref. [2]), even when $C_0^+$ (and hence $\sigma$) approaches zero. This is relevant here, as we consider modes for which the infernal drive is much weaker than the interchange drive. We replace $r_1$ with $r_\star$, and $\lambda$ by $\lambda_0$ (eq. (8), we only need interchange terms) in (5) and obtain:

$$\xi_\gamma^{(m)}(r) = \frac{\sigma}{\lambda_0}\left(\frac{r_\star^m J_m\left[k(\gamma/\omega_A)r\right]}{rJ_m\left[k(\gamma/\omega_A)r_\star\right]} - r^{m-1}\right). \tag{14}$$

This approximation, instead of the Heaviside one, will be injected in the expression for $\delta W$ (a general validity check is performed in Appendix C). Since this ansatz $\xi_\gamma(r)$ now depends on the growth rate $\gamma/\omega_A$, the potential energy $\delta W$ will itself depend on the growth rate.

Solving equation (13) then gives the dispersion relation we sought, without restricting to the most unstable mode. Note that the most unstable mode in (14) recovers $\xi(r) = \xi_0\left(r^{m+1}/r_\star^{m+1} - r^{m-1}/r_\star^{m-1}\right)$ which is the result obtained for infernal modes with weak shear.

Keeping the variation in the $q$ profile, and general $m$, requires treatment of the term in $\delta W$: $(1-m)\frac{r^{m+1}}{r_\star^{m+1}}\xi^{(m)}\frac{\mathrm{d}}{\mathrm{d}r}\left(\left(\frac{1}{q}-\frac{1}{q_s}\right)^2\right)$. We thus need an explicit expression for $\left(\frac{1}{q}-\frac{1}{q_s}\right)$. To keep things tractable, we approximate $q$ as linear between the beginning of the low-shear zone and $r_\star$. We define:

$$C_q = \frac{q_\star - q(r_0)}{r_\star - r_0}\,, \tag{15}$$

where $r_0$ marks the beginning of the low shear zone, for the applications in this paper: $r_0 = 0$. We replace $\xi^{(m)}$ by the ansatz (14), and perform to integrals needed in eq. (12), to obtain:

$$\delta W_\gamma = \frac{(1-m)}{2(1+m)^2}\frac{r_\star^2(\alpha/r)^2 J_{2+m}[k(\gamma/\omega_A)r_\star]C_q^2\Lambda_{m,n}}{q_s^6\lambda_0 J_m[k(\gamma/\omega_A)r_\star]} - \frac{(\alpha/r)^2 J_{m+2}[k(\gamma/\omega_A)r_\star]\Lambda_{m,n}\widehat{D_I}}{4(1+m)^2\lambda_0 J_m[k(\gamma/\omega_A)r_\star]q_s^4}\,. \tag{16}$$

We again replaced $\sigma$ in eq. (16) using its definition in terms of the quantity $\Lambda_{m,n}$ (given in Eq. (6)), which can be expressed analytically for this case as: $\sigma = \frac{\alpha}{2\,r\,q_s^2}C_0^+ \approx \left(\frac{\alpha}{r}\right)^2\Lambda_{m,n}\xi_0\frac{r_\star^{3+m}}{2+2m}\frac{1}{q_s^2 r_\star^{2+2m}}$. To arrive at this form, we approximated the integral $\left(\int_0^{r_\star} r^{2+m}\xi\, dr\right)$ by assuming the shape of the most unstable mode (Heaviside ansatz). This approximation will thus be less accurate for the higher oscillating modes.

### 3. *Final dispersion relation with magnetic shear*

The dispersion relation with shear is then the combination of equations (13) and (16):

$$\begin{aligned} D(\gamma^2/\omega_A^2) =& \frac{(1-m)}{2(1+m)^2}\frac{r_\star^2(\alpha/r)^2 J_{2+m}[k(\gamma/\omega_A)r_\star]C_q^2\Lambda_{m,n}}{q_s^6\lambda_0 J_m[k(\gamma/\omega_A)r_\star]} - \frac{(\alpha/r)^2 J_{m+2}[k(\gamma/\omega_A)r_\star]\Lambda_{m,n}\widehat{D_I}}{4(1+m)^2\lambda_0 J_m[k(\gamma/\omega_A)r_\star]q_s^4} \\ &+ \frac{1}{\pi}\sqrt{\frac{\gamma^2(1+2q_s^2)}{m^2\omega_A^2} + \frac{\Delta q_\star^2}{q_\star^2 q_s^2}}\sqrt{\frac{s_\star^2}{q_\star^2} + \frac{\Delta q_\star q_\star'' r_\star^2}{q_s^2 q_\star^2} - \frac{2\Delta q_\star s_\star^2}{q_s^2 q_\star}} = 0\,, \\ \text{with}\quad & k(\gamma/\omega_A) = \sqrt{\lambda_0\left(\frac{\gamma^2(1+2q_s^2)}{m^2\omega_A^2} + \frac{\Delta q^2}{q_s^4}\right)^{-1}}\,. \end{aligned} \tag{17}$$

For convenience, we recall here the constants needed in the root-finding process of solving (17) for $\gamma/\omega_A$:

- $\Lambda_{m,n}$ is associated with the sideband perturbation and can be approximated analytically as $\Lambda_{m,n} = \frac{(m+1)(m+2)}{2m(2m+1)}\left(\frac{q_a}{q_s} - 1\right)\left(\frac{r_\star}{1-r_\star}\right)$.
- $\lambda_0$ is given by equation (8) (this is an approximation, since this definition of $\lambda$ was obtained assuming a shearless profile, and should feature more terms with shear. We choose to keep a "simple" equation, and as will be seen, this is enough to retrieve correct growth rates).
- $r_\star$ can indicate $r_\star = r_s$ for a monotonic $q$ profile, or $r_\star = r_{min}$, where $q(r_{min}) = q_{min}$ for a reversed $q$ profile. $\Delta q_\star$ is then defined as $\Delta q_\star \equiv q(r_\star) - q_s$, so $\Delta q_\star = 0$ for a monotonic $q$ profile. Following the same logic, $s_\star$ will be zero for a reversed $q$ profile.
- $C_q$ characterizes the shear of the $q$ profile in the low shear zone and is given by (15).
- $\kappa$ is the elongation, and $\delta$ is the triangularity. Both $\kappa$ and $\delta/r$ are considered constant.

***Growth rates and eigenfunctions solutions*** Figures 4 and 5 show results for a slightly reversed, non resonant $q$ profile, and elongated flux surfaces. Although these modes are interchange modes, the low shear region allows a full spectrum of unstable modes to emerge (as seen here using the modular solver, *i.e.* by solving the full equations). The corresponding plasma displacements of the four most unstable $m = n = 1$ modes are shown in Fig. 4a. We observe that the eigenfunctions have a

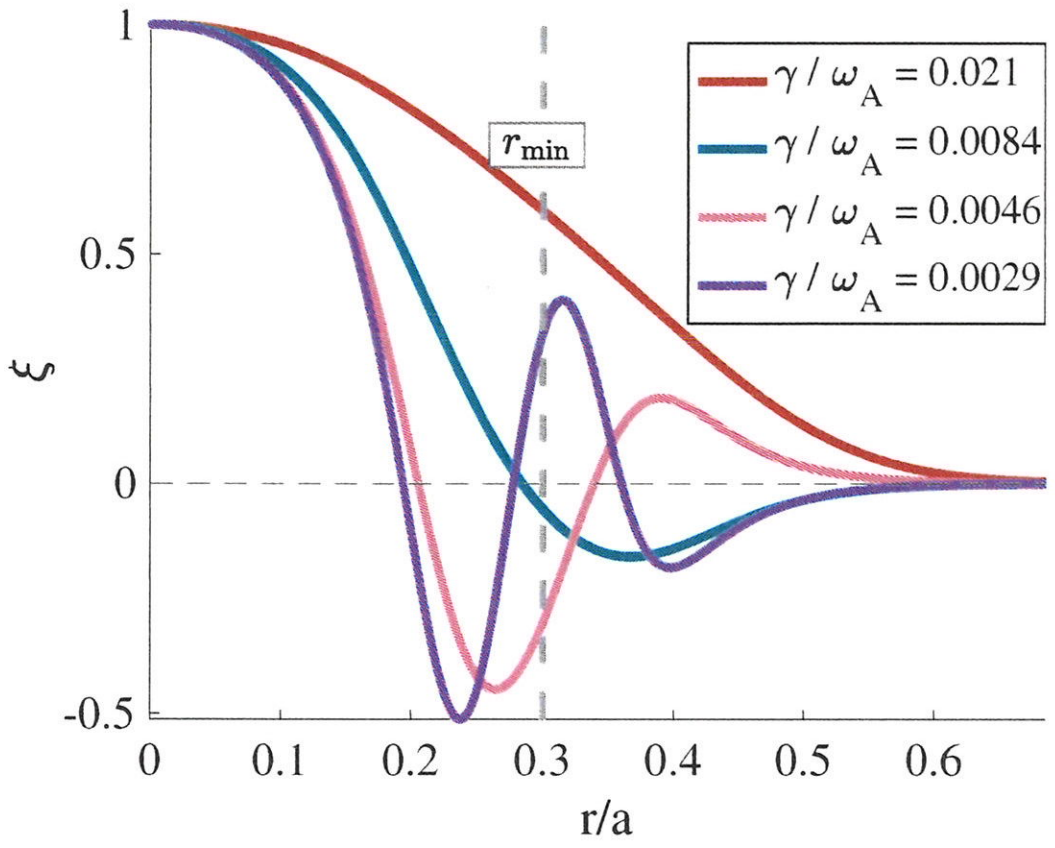

(a) Modular solver results : radial plasma displacements for the first 4 modes, and their associated growth rates.

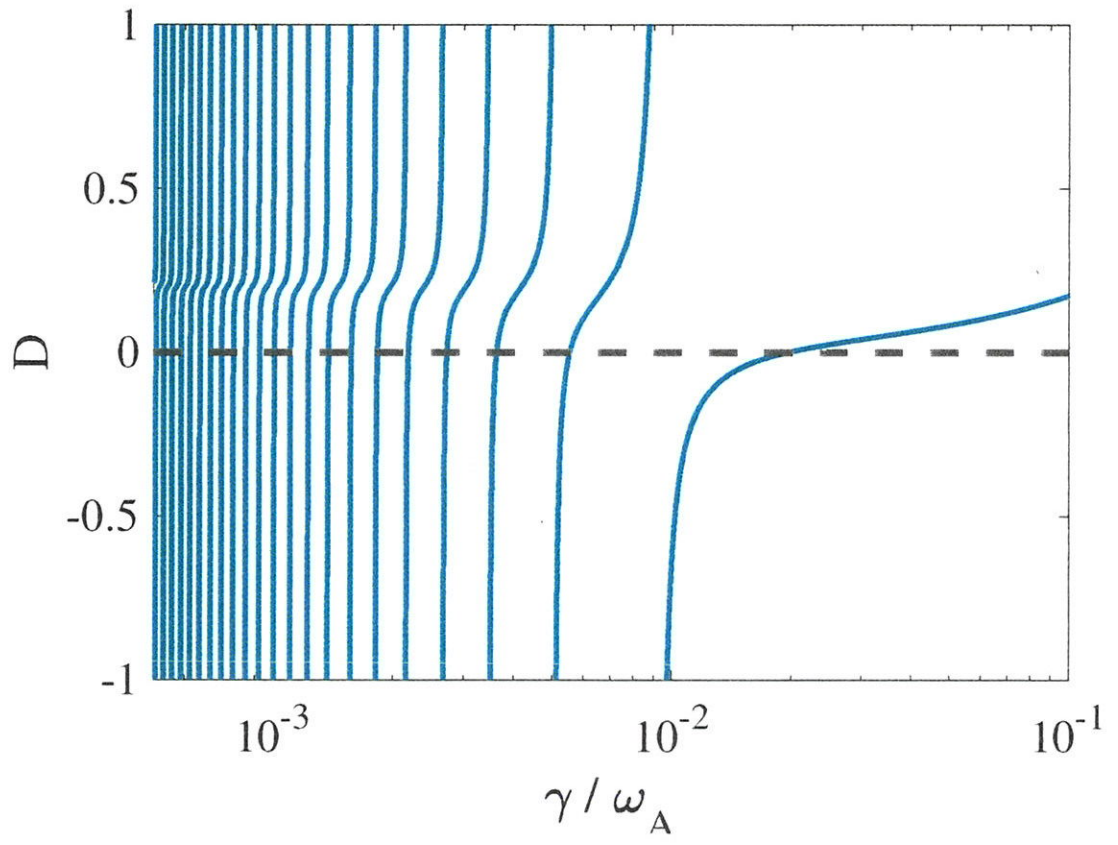

(b) Analytic description : dispersion relation with shear $D(\gamma^2/\omega_A^2)$ (equation (17)).

Figure 4: $m = n = 1$ spectrum obtained with the profiles of Fig. 5 ($\kappa = 1.5$, $\delta_1 = 0$): comparison of the full equations solutions (top panel) against the dispersion relation (17) for the growth rates (bottom panel).

slightly more complicated structure than those obtained for a monotonic $q$ profile (see Figure 3). We attempt to recover these growth rates using the analytical dispersion relations. When solving the dispersion relation with shear that we just derived, we are searching for the roots of the function given in (17). This function is showed in 4b for the case at hand: the $q$ profile shown on the left in Fig. 5, which produced the $m = n = 1$ spectrum found with the modular solver.

The roots of this function give the growth rates, and are showed on the middle plot of Fig. 5. This figure shows both the $m = n = 1$ and $m = n = 2$ spectra, comparing the dispersion relation with shear that we just derived, and the expression without shear given in (3). We expect the latter to be less accurate, since the $q$ profile is now sheared. We indeed see that the sheared dispersion relation roots are closer to the solver's results, better capturing the growth rate, especially for the more unstable modes. This good agreement has been found for a wide range of values of $\alpha$. For less unstable modes, the agreement is slightly poorer, which is expected due to the approximation made for the amplitude factor $\sigma$, primarily valid for the most unstable mode. The dispersion relation 17 is thus a generalisation of the standard interchange dispersion relation to reverse shear cases, and additionally captures the complete spectrum of modes emerging in low shear regions.

### *4. Most unstable mode & stability criteria for reverse shear cases*

An approximate expression for the most unstable mode growth rate in reverse shear configurations can be derived from the dispersion relation (13) by evaluating $\delta W$ with an ansatz for $\xi$ adapted to the most unstable mode in non-resonant profiles, namely $\xi = \xi_0 \left(\frac{r}{r_\star}\right)^{m-1} \left(1 - \left(\frac{r}{r_\star}\right)\right)^2$. This yields the growth rate for the interchange mode in reverse magnetic shear configurations:

$$\gamma^2/\omega_A^2 = \frac{m^2}{1+2q_s^2}\left[\frac{\pi^2\left(\widehat{D_I} q_s^2 + 2(m-1) r_\star^2 C_q^2\right)^2}{4(1+m)^2(2+m)^2(3+2m)^2 r_\star^2 q_s^4 q_*'' \Delta q_*} - \frac{\Delta q_*^2}{q_s^4}\right]. \tag{18}$$

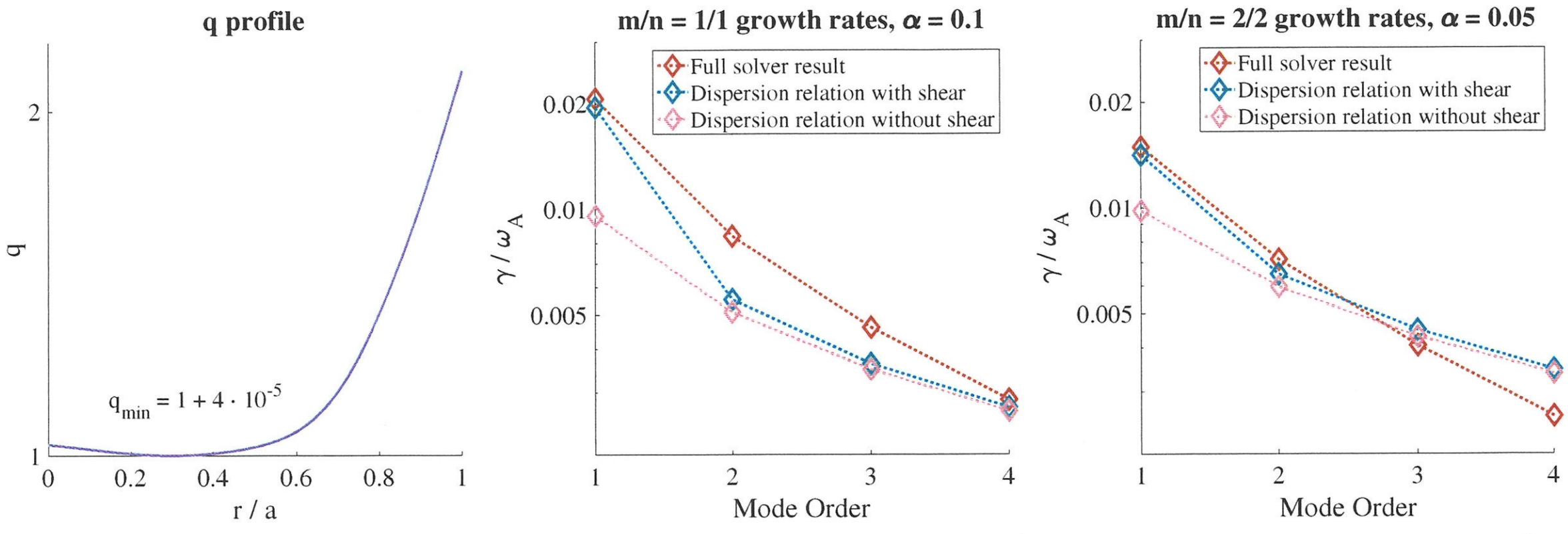

Figure 5: Reversed $q$ profile giving rise to a spectrum of infernal modes, for a parabolic pressure profile with $\alpha(0.3) = 0.1$ or $0.05$, $\kappa = 1.5, \delta_1 = 0$. The growth rates of the modes are obtained solving the full equations with the modular solver. These values are compared to solving the analytic dispersion relations, with shear (eq. (17)), or without (eq. (3)).

From this, the instability condition for the most unstable mode follows:

Interchange instability criterion for reverse shear cases:

$$\frac{\Delta q_\star^2}{q_s^4} 4(1+m)^2(2+m)^2(3+2m)^2 < \frac{\pi^2}{r_\star^2 q_s^4 q_\star'' \Delta q_\star} \left(\widehat{D_I} q_s^2 + 2(m-1) r_\star^2 C_q^2\right)^2 . \tag{19}$$

As with the no-shear criterion (10), the left-hand side is the stabilising field line bending term, while the right-hand side contains the destabilising contributions (if $\widehat{D_I} > 0$). We can also reformulate this criterion analogously to the Mercier criterion, if $\widehat{D_I} < 0$ is stabilising:

Interchange instability criterion for reverse shear cases, if $\widehat{D_I} < 0$ :

$$\left(-\widehat{D_I}\right) + \frac{2(1+m)(2+m)(3+2m)}{\pi} r_\star \frac{\Delta q_\star^{3/2}}{q_s^2} \sqrt{q_\star''} < \frac{2(m-1) C_q^2 r_\star^2}{q_s^2} . \tag{20}$$

This inequality provides a criterion for the onset of interchange instability in reverse shear configurations, including shaping effects.

Overall, we find that non-resonant $q$ profiles give rise to rich physics, with the modular solver revealing potentially infinite spectra of unstable infernal modes for single mode numbers $m$ and $n$. This variety of instabilities is even more pronounced with the addition of shaping of the flux surfaces. This motivated us to derive analytic expressions for these spectra, aiming to predict whether a spectrum appears and to estimate the corresponding growth rates, without the need to solve the full equations with the modular solver. We first include shaping effects into the dispersion relation for monotonic $q$ profiles. This allows us to observe how shaping, through modification of the Mercier term, can completely change the nature of the spectrum, while having a relatively weak impact on the most unstable growth rate. We find that even a moderate amount of shaping can transition the system from a single unstable root to numerous different modes. We then extend the study to sheared profiles, which require retaining additional terms. Adopting an energy minimization method, we derive a fairly simple dispersion relation valid for interchange unstable cases. Even though several approximations were made (using a shearless ansatz and calculation for $\lambda$, and keeping only interchange terms in the potential energy), the resulting growth rates agree well with the results of the modular solver. For each of these cases, we obtain modified Mercier criteria as a convenient way to quickly assess the stability of a given configuration.

## III. RESISTIVE MODES AND SPECTRA IN RESONANT SCENARIOS

### A. Combination of negative triangularity and resistivity

The previous section explored non-resonant scenarios, monotonic or reversed, which are relevant to a wide range of advanced tokamak regimes. When avoiding rational surfaces, ideal MHD is a sufficient description. However, in practice, most advanced scenarios in tokamaks will encounter rational surfaces with higher $q_s$ values, such as the $m/n = 2/1$, which give rise to the usual tearing mode problem. Even in the core of these advanced scenarios, it is still interesting to consider $q$ profiles with a $q_s = 1$ resonance. Hybrid $q$ profiles with $q_s = 1$ might fail to avoid the rational surface (since measurements in the core are difficult, it is hard to know the exact $q$ profile in practice). Advanced inductive tokamak scenarios, which have a lower bootstrap current fraction, are also likely to encounter a $q_s = 1$ resonance. With resonance, resistivity becomes crucial, and its effects must be taken into account to avoid overlooking important modes. To explore the combined effects of shaping, infernal physics, and resistivity diffusion, we make use of our resistive modular solver [2], and the classical Glasser–Greene– Johnson (GGJ) analytic dispersion relations [18].

With resonance, the spectrum behaviour we presented in section II changes. Except for $q_s < 1$, where the average curvature is unfavourable, there is no spectrum of modes for resonant $q$ profiles, no matter the pressure or resistivity drive when the flux surfaces are circular. But if another instability drive comes into play, namely unfavourable shaping of the flux surfaces (*i.e.* positive elongation with no or negative triangularity), things change. Looking at the Mercier criterion (1), it is easy to see how the destabilizing effect from shaping can produce a spectrum via causing unfavourable curvature, even for $q_s = 1$ or $q_s > 1$.

#### 1. *Impact of negative triangularity on tearing modes*

It is interesting to linger a bit more on the topic of negative triangularity (NT). This shaping of the flux surface is indeed another envisioned solution for future reactors, since it is promising for reducing fluctuation levels in the edge and for minimising ELMs. For the maturity of such plasma confinement concepts, we need more experimental and theoretical studies on *e.g.* tearing mode stability for NT [19]. Usually tearing modes are stabilised by pressure. But this is necessarily true with unfavourable shaping (again because of this similar effect between shaping and curvature which can clearly be seen from the corrected Mercier criterion (1)). This is illustrated in Figure 6 via calculation with the modular solver. We consider a classical case of a hybrid $q$ profile, shown in black in Figure 7, which has a resonance on the 2/1 surface, typically around $r = 0.8$. Without any shaping or with positive triangularity, the tearing mode is completely stabilised for a sufficient pressure (in this case around $\alpha \sim 0.2$). This pressure stabilisation is however lost with negative triangularity. Even for the moderate case we are looking at ($\kappa = 1.5$ and $\delta_a = -0.3$), the growth rate diminishes at a much lower rate, instead of vanishing with increasing $\alpha$. This effect, as will be seen later, can be understood from equations (27) and (28).

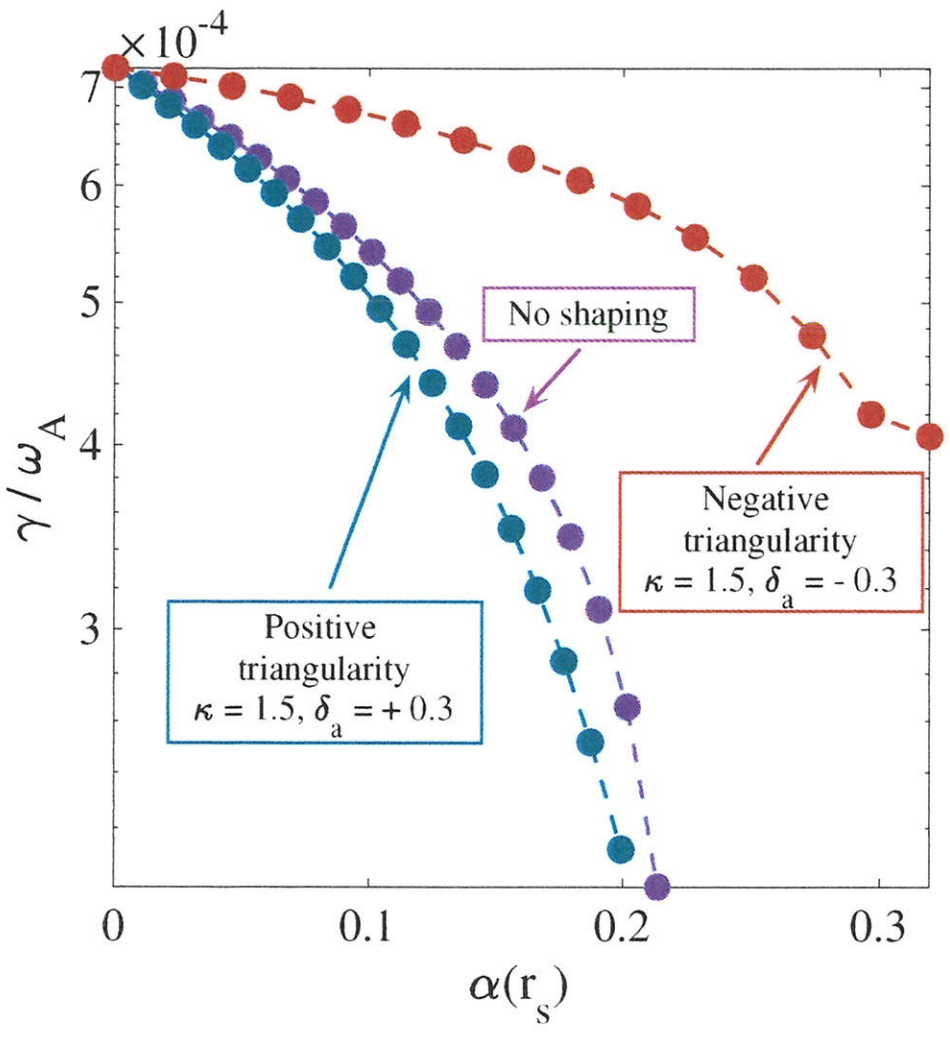

Figure 6: Linear growth rates of the $m/n = 2/1$ tearing mode with respect to pressure (parabolic pressure profile), for the hybrid $q$ profile shown in black in Fig. 7 (with $r_s = 0.75$, $q_0 = 1.2$, and $q_a = 4$), $S_L = 10^6$.

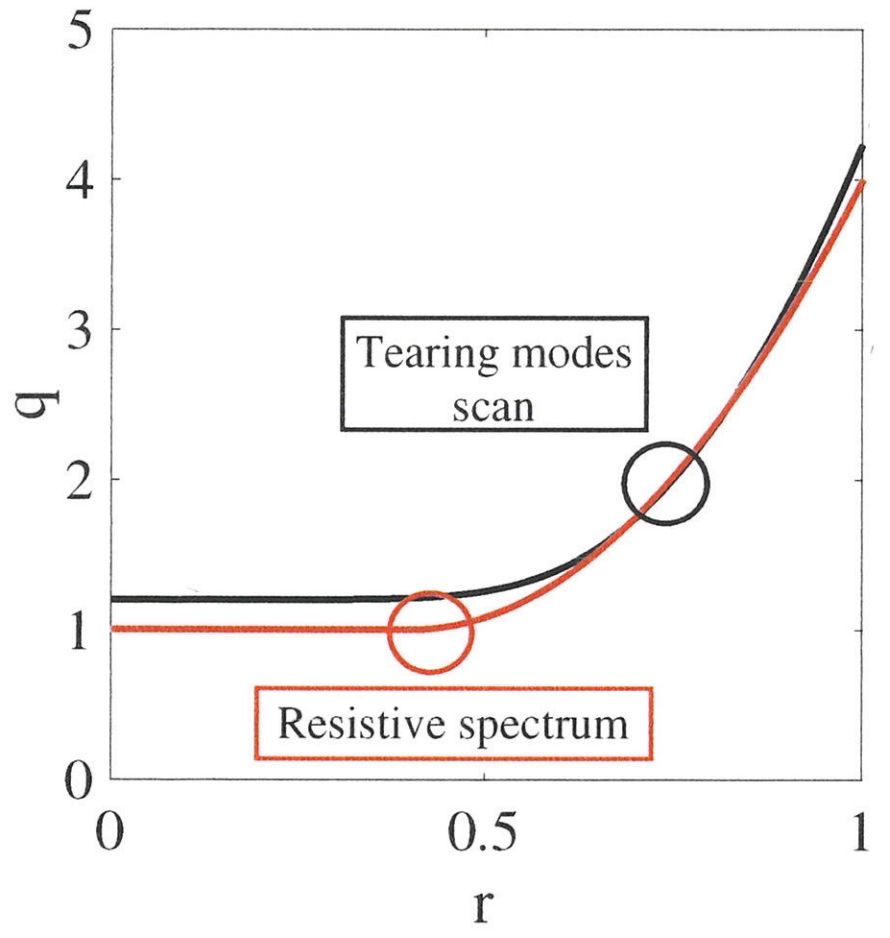

Figure 7: $q$ profiles for advanced scenarios. Rational surfaces are circled, yielding either tearing modes (Fig. 6) in the black configuration or a spectrum of resistive modes (Fig. 8) in the red configuration.

### 2. *Emergence of a resistive internal kink spectrum at the $q = 1$ surface*

Thus, negative triangularity leads to a loss of pressure stabilisation for tearing modes. But what about negative triangularity and the spectra of instabilities in the core ? With unfavourable shaping, we have ideal and resistive spectra of instabilities even for a resonant $q$ profile for $q_s = 1$. An interesting feature of the ideal spectrum is that it completely disappears if we lower the pressure enough, even though there is still an ideal unstable eigenmode, with non negligible growth rate ( $\gamma_{ideal}/\omega_A = 9.7{\cdot}10^{-4}$ for $n = 2, m = 2$ in Figure 8). This is a similar phenomenon to what we describe in [2], for $q_s < 1$ there, without shaping. Nevertheless, if there were to be finite resistive diffusion for this same pressure, a resistive spectrum would appear. An example of this phenomenon is shown in Figure 8, for the $q$ profile in red in Figure 7. It displays a resistive spectrum, since with this pressure, shaping and $q$ profile, we would have only one ideal mode (with $\gamma_{ideal}/\omega_A = 9.7{\cdot}10^{-4}$). We emphasise that this spectrum would not be there without shaping. Here negative triangularity gives rise to much more exotic physics in the plasma core.

The plasma displacements shown in Figure 8 display typical infernal features, with broad structures extending across the entire low-shear region, inside the rational surface at $r = r_s$. As for the ideal spectrum, the less unstable modes exhibit increasingly oscillatory behaviour. A new feature however emerges in the resistive layer around the rational surface: the eigenfunctions become more irregular and develop small-scale structures, which is typical of resistive instabilities. This confirms that we are indeed observing a spectrum of resistive infernal modes, driven unstable by negative triangularity. Figure 9 plots the perturbation to the magnetic flux associated with this spectrum, in the radial direction. The perturbations associated with the higher oscillating modes are also shown. We can see that for all modes in the spectrum, there is a non zero radial perturbation of the magnetic flux at the rational surface. This means there is magnetic reconnection when the spectrum appears, and these modes collectively could contribute to eventual stochastization of the core, an extension to a non-linear simulation.

## B. Resonant reversed shear scenarios

Let us now examine the stability of resonant reversed shear scenarios, making use of the modular solver. We compare two types of reversed shear $q$ profiles, shown in Figure 10: one with a strong shear reversal, and another almost flattened around the rational surface, at $q_s = 1$. The dependence of the growth rates on the poloidal mode number $m$ is first studied. We extend previous work in this area, since this has been studied for double tearing modes (e.g. in [20]), but remains less explored for resistive infernal modes, especially with a full-MHD mode, and leading order shaping effects included. The results are presented in Figure 11, for a low-pressure case, and different shaping configurations. For all cases, the $m = 1$ mode is an ideal mode, but for the other poloidal mode numbers, the modes would be stable in an ideal simulation. For positive triangularity, and circular flux surfaces (left and middle plots), low-$m$ modes are dominant, with the $m = 1$ mode being the most unstable. This holds true for both the strong and weak shear reversal profiles. With negative triangularity however, even if the $m = 1$ mode remains the most unstable overall, the trend with $m$ changes. For the weak shear reversal profile, the growth rates increase steadily with $m$, eventually exceeding those of the strong reversal profile at sufficiently high $m$. This dynamic may be linked to the greater proximity of the weak shear reversal profile to the $q_s = 1$ surface, making it more sensitive to the destabilizing effect of negative triangularity at higher mode numbers. Overall, we find that for small pressure, the strong shear reversal scenario is more unstable than the flatter one, across most of the $m$ spectrum. There is an exception for negative triangularity cases and high $m \sim 10$, where the weak shear reversal profile is more unstable, due to enhanced sensitivity in the vicinity of $q_s = 1$.

Now what are these modes exactly? To gain insight, we begin by examining their dependence on pressure gradients. Figure 12 shows for the same profiles the evolution of the growth rates for the $m = 1$, $m = 2$, and $m = 3$ modes, for the case with circular flux surfaces. For both profiles (weak and strong shear reversal), the growth rates increase with $\alpha$, indicating that these modes are not purely (double) current driven tearing modes, therefore these are at least partially pressure-driven.

To further investigate how the modes evolve with increasing pressure, we examine the structure of the eigenfunctions. For $m > 1$ modes, the shape of the plasma displacement becomes broader as we increase $\alpha$, progressively taking on the characteristic shape of an infernal mode. This is the case for the weakly sheared $q$ profile, and can be seen on Figure 13. For the $q$ profile which has a higher shear throughout, this is not so much the case, and the eigenfunctions retains a structure typical of a double tearing mode ($\xi$ is odd in $r$ across the rational surfaces), as shown in Fig. 14, regardless of the pressure gradient. Indeed, for this profile, the higher magnetic shear across the radial direction prevents the development of the broad infernal-like structure, and the more localized tearing-like structure persists.
For the $m = n = 1$ mode (which is an ideal mode), the behaviour is quite similar between the weakly and strongly reversed shear profiles. As pressure increases, the importance of the second rational surface becomes more pronounced. This evolution of the eigenfunction is shown in Fig. 15, as an illustration valid for both profiles.

In summary, for very small pressure we find that no matter the shaping of the flux surfaces, the scenario with strong shear reversal is more unstable than the weakly sheared one, for all mode numbers $m$. The only exception is with negative triangularity: for sufficiently high $m$, growth rates of the weakly sheared profile increase again and exceed those of the strong shear profile. This is likely due to the wide region with small field line bending near $q_s = 1$ present in the low shear scenario. It is more sensitive to variations in the magnetic well which can be cause by pressure or shaping of the flux surfaces. For the same reason, as the pressure increases, the low shear scenario becomes overall much more unstable. While at low pressure all resistive modes initially display a typical double tearing mode structure, we find that they are destabilized by increasing pressure. This, and the change in eigenmode structure, especially for the scenario displaying a weak shear, indicates how the double tearing mode branch transforms into the resistive infernal branch as pressure increases. These resonant reversed shear configurations thus display a rich variety of behaviour, strongly depending on the pressure gradient.

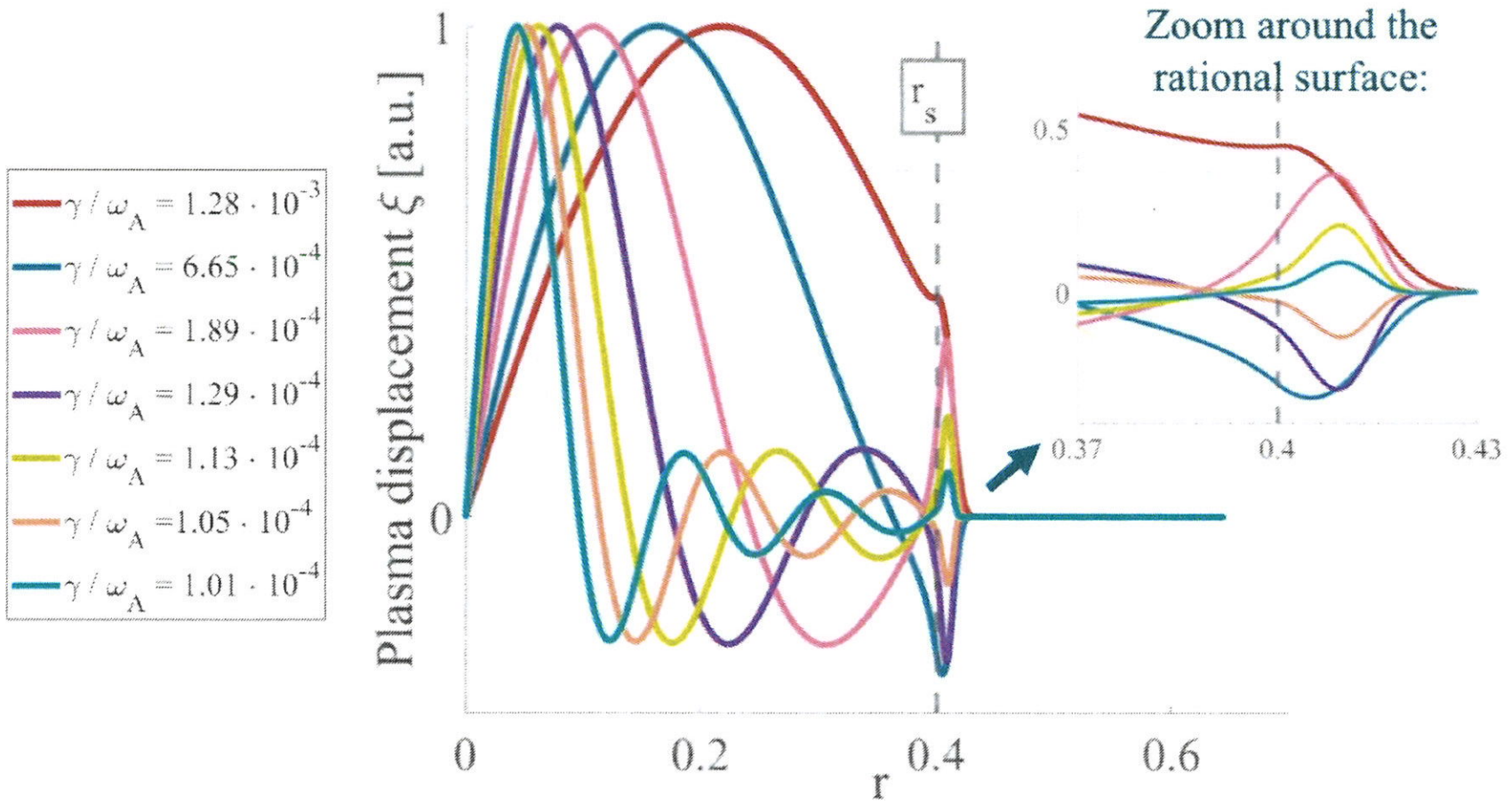

Figure 8: Spectrum of resistive $m/n = 2/2$ mode in NT, advanced inductive configuration (red $q$ profile in Fig. 7, $\kappa = 1.3, \delta_a = -0.1$), $\gamma_{\text{ideal}}/\omega_A = 9.7 \cdot 10^{-4}$, $\alpha(r_s) = 2 \cdot 10^{-3}$ (or $\beta_p(r_s) = 3.75 \cdot 10^{-3}$, parabolic pressure profile), $S_L = 10^6$.

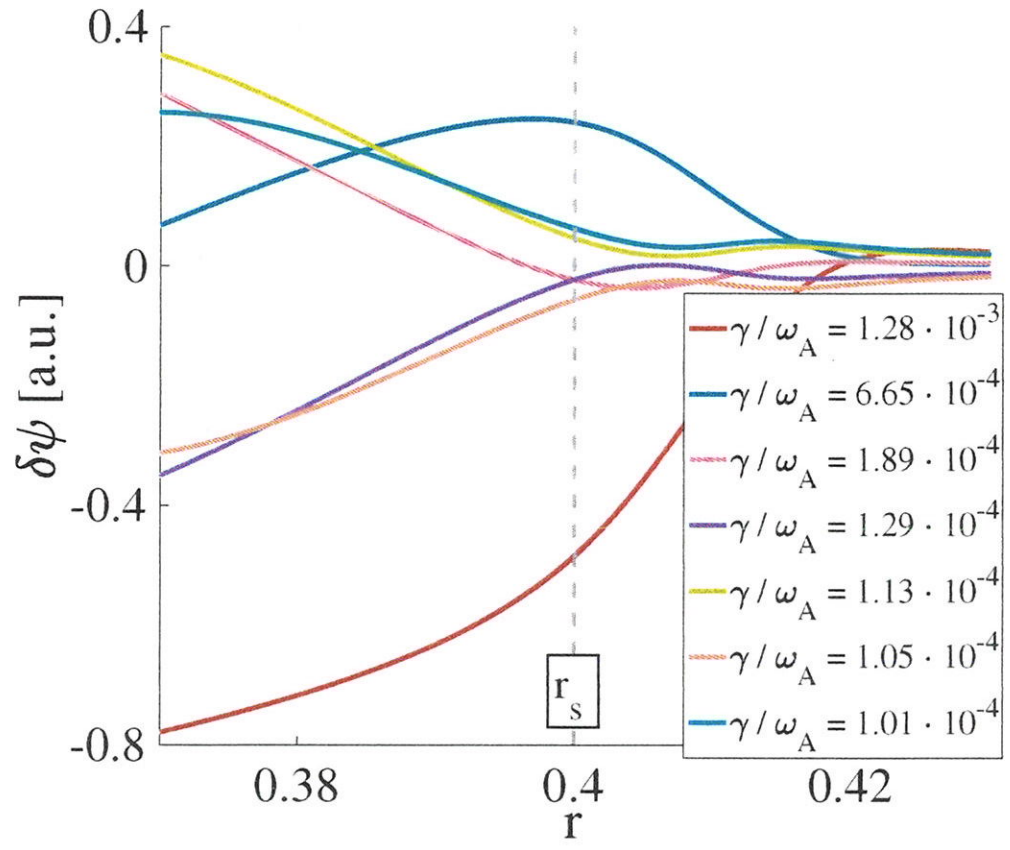

Figure 9: Radial perturbation of magnetic flux associated with the $m/n = 2/2$ spectrum shown in Fig. 8: zoom around the rational surface at $r = r_s$.

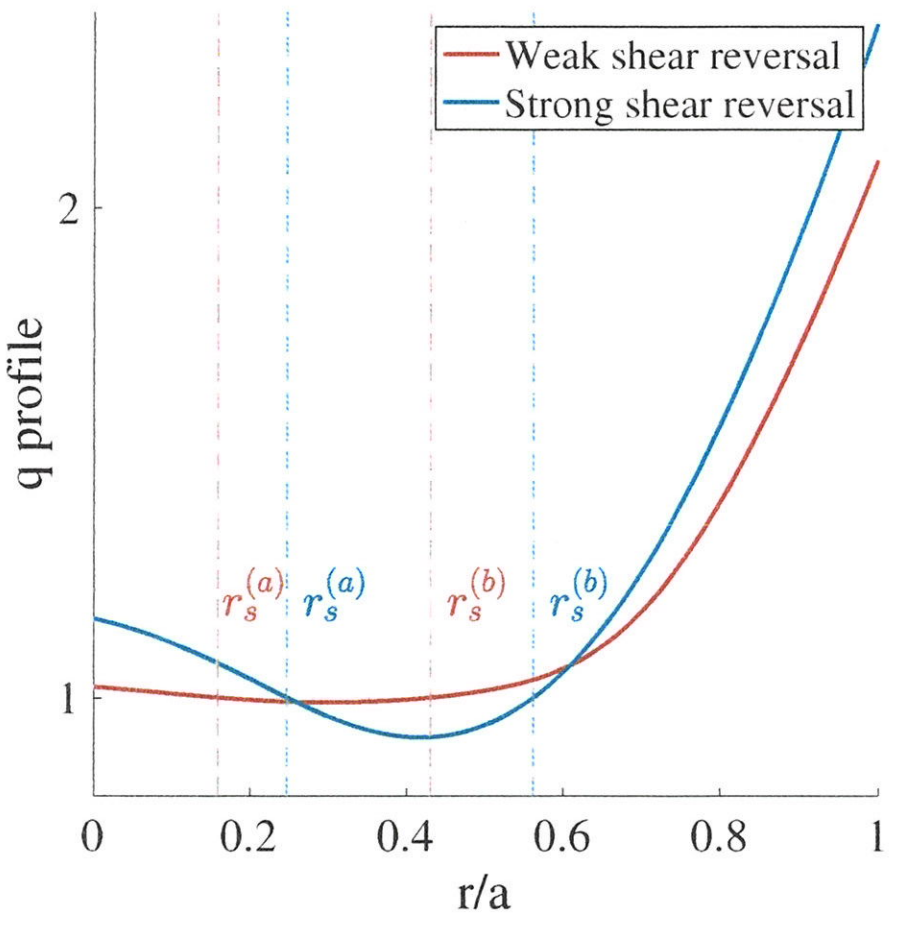

Figure 10: Two types of reversed $q$ profiles: comparing strong and weak shear reversal.

## IV. IMPORTANT CONSIDERATIONS FOR ACCURATE NUMERICAL OR ANALYTICAL DESCRIPTIONS

In the previous sections, we explored the richness of modes arising in low-shear configurations. However, the ability to resolve these structures can strongly depend on the assumptions made in numerical codes. Certain simplifications may filter out some of these modes, potentially overlooking important dynamical behaviours. This is what we explore in the next section.

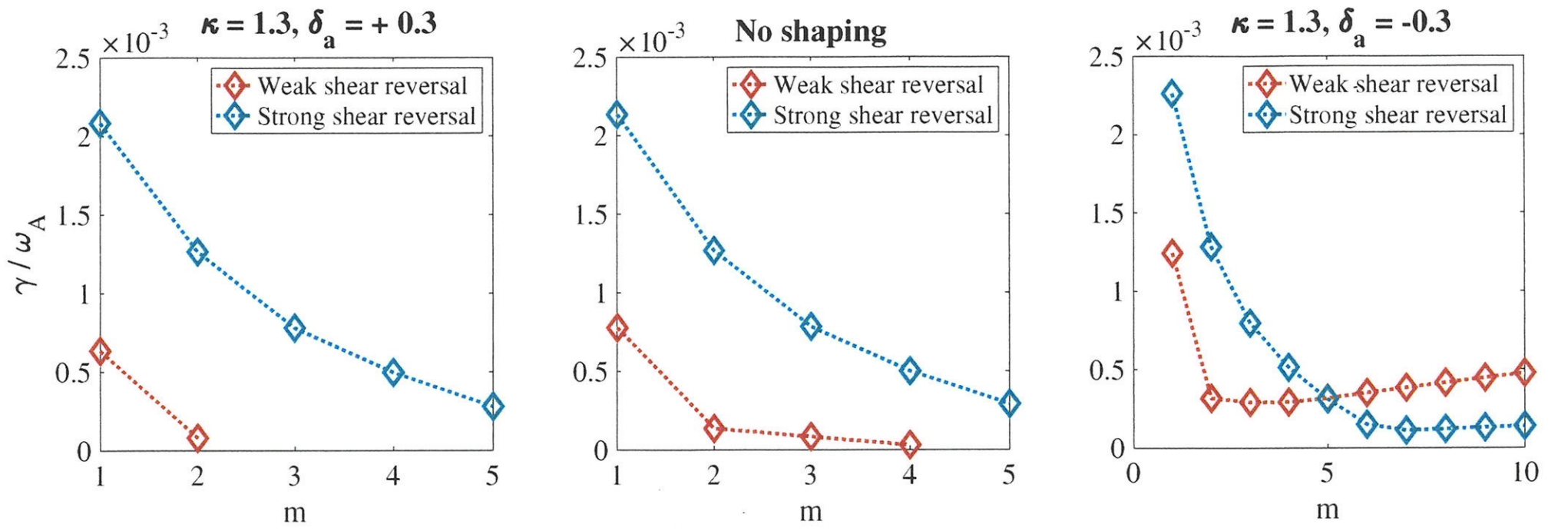

Figure 11: $n = m$ resistive modes growth rates for the two reversed shear scenarios shown in Fig. 10, with $q_s = 1$. Different shaping configurations are compared, for a parabolic pressure profile with $\alpha(r_s^{(a)}) = 0.001$. $S_L = 10^6$.

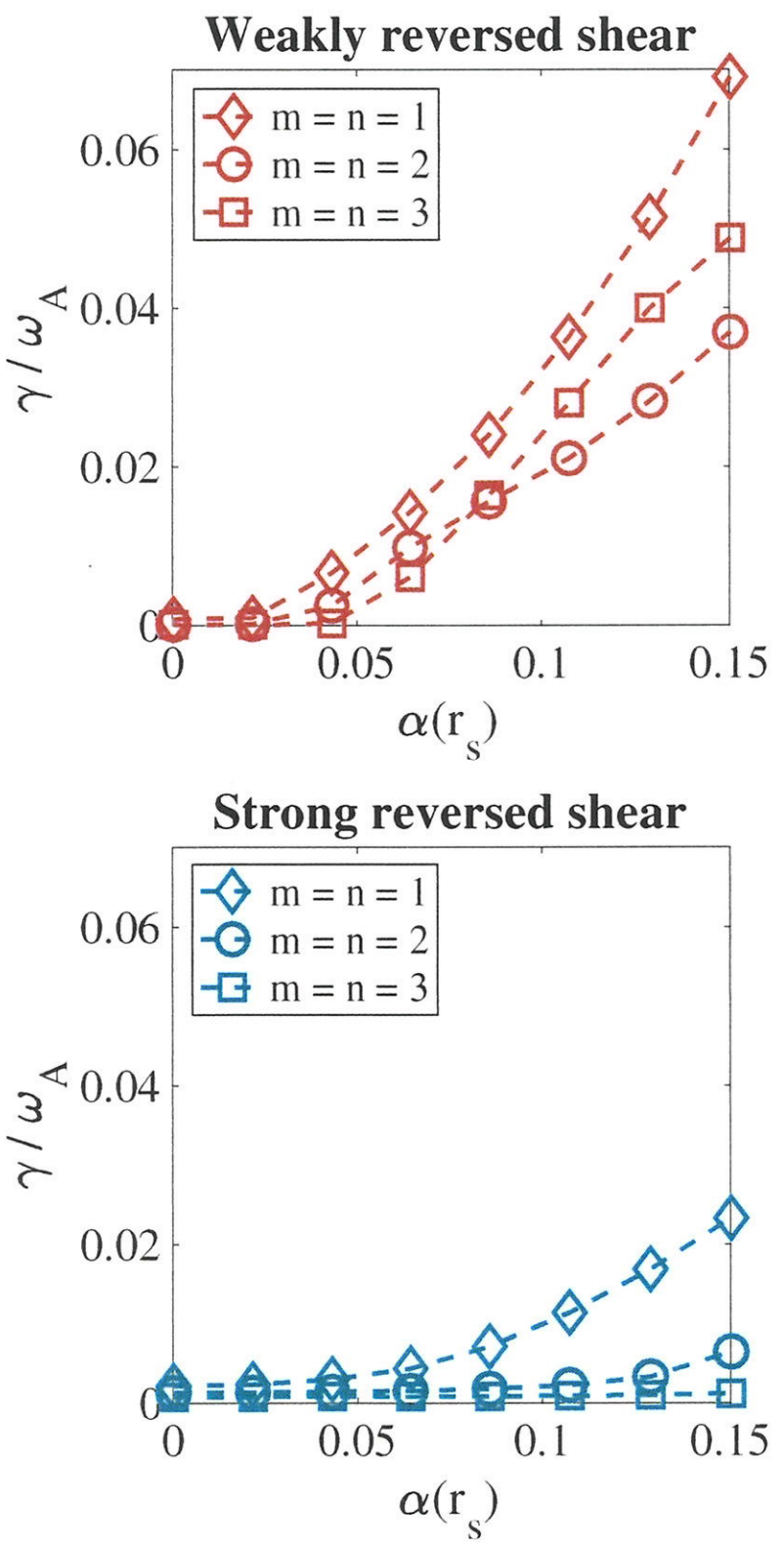

Figure 12: Growth rate evolution with increasing pressure (parabolic pressure profile) for the two reversed shear scenarios shown in Fig. 10, with $S_L = 10^6$, and circular flux surfaces.

## A. Impact of neglecting toroidal effects on tearing mode growth rates

In the absence of neoclassical effects, the tearing parameter $\Delta'_R$ (first introduced in Ref. [21]) determines tearing mode stability, with $\Delta'_R < 0$ ensuring the mode is stable. The linear growth rate of tearing modes can also be derived from this parameter. In this section, we examine whether including toroidal effects at different stages of the growth rate calculation significantly impacts the result, in order to identify which aspects in numerical codes are essential to obtain correct growth rates. This quantity $\Delta'_R$ can be understood by considering a thin resistive layer of width $\delta_\eta$, around the rational surface where $q = m/n$, where field line bending becomes negligible while inertia and resistivity are important. Outside the layer, in the two exterior regions, field line bending dominates and inertialess ideal MHD applies, allowing the use of simple cylindrical equations. For the perturbed magnetic flux $\psi$ this gives:

$$\begin{aligned} 0 =& r^2\psi'' + r\psi' - m^2\psi + r\frac{mq}{nq-m}\frac{R_0}{B_0}\frac{\mathrm{d}}{\mathrm{d}r}(J_\phi(r))\psi \\ &+ \frac{m^2q^2}{(nq-m)^2}\frac{s(r_s)^2}{q_s^2}D_I(r_s)\psi\,, \end{aligned} \tag{21}$$

where the last term is an additional contribution compared to the usual cylindrical equations. It corresponds to finite pressure (toroidal) effects, and we choose to include it to assess its importance in the calculation of the tearing parameter. $D_I(r)$ is the contribution from the toroidal magnetic well and the shaping of the flux surfaces, given in (1). We repeat it here for convenience:

$$D_I(r) = \frac{\epsilon\alpha}{s(r)^2}\left[\frac{1}{q_s^2} - 1 + \frac{3}{4}(\kappa-1)\left(1 - 2\frac{\delta}{\epsilon}\right)\right]. \tag{22}$$

This equation (21) can be solved in the two exterior regions without much difficulty, at least in the absence of the toroidal term $D_I(r_s)$. The matching of the two outer solutions across the resistive layer is done by using the constant $\psi$ approximation: setting $\psi(r_s - \delta_\eta) = \psi(r_s + \delta_\eta)$. Indeed, the perturbed flux varies only weakly in the layer, which is very thin ($\delta_\eta \ll 1$). However the exterior solutions approach $r_s$ differently from the inside and outside the rational surface (indicating a current sheet in the layer), and $\Delta'_R$ characterizes the discontinuity in the slope of $\psi$:

$$\Delta'_R = \lim_{\delta_\eta \to 0} \frac{\delta_\eta \psi'}{\delta_\eta \psi} \Bigg|_{r_s - \delta_\eta r_s}^{r_s + \delta_\eta r_s} . \tag{23}$$

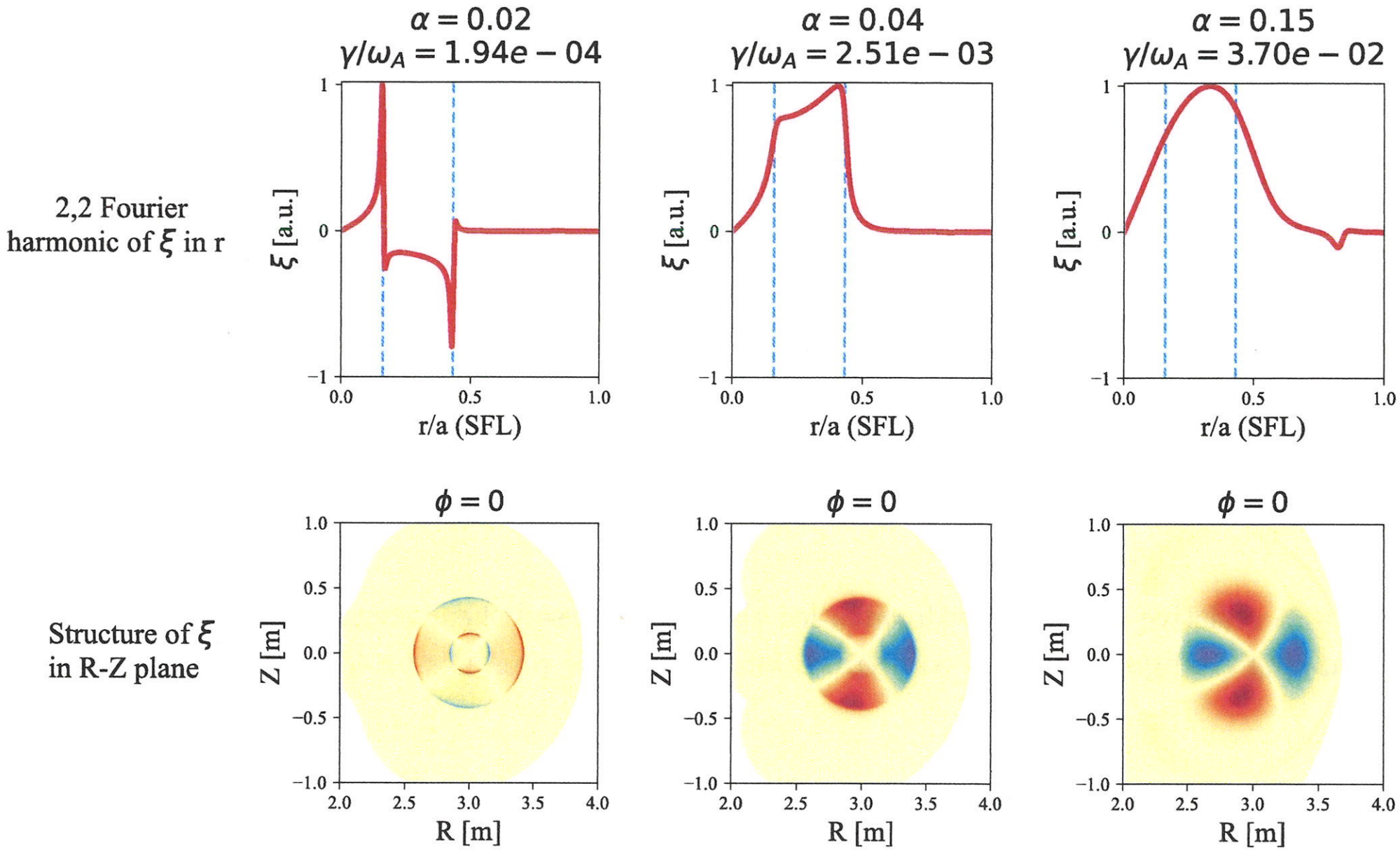

Figure 13: Radial plasma displacement structure and contour plots of the $m = n = 2$ mode for different pressure drives, in the weakly reversed shear profile. Parabolic pressure profile and circular flux surfaces; $S_L = 10^6$.

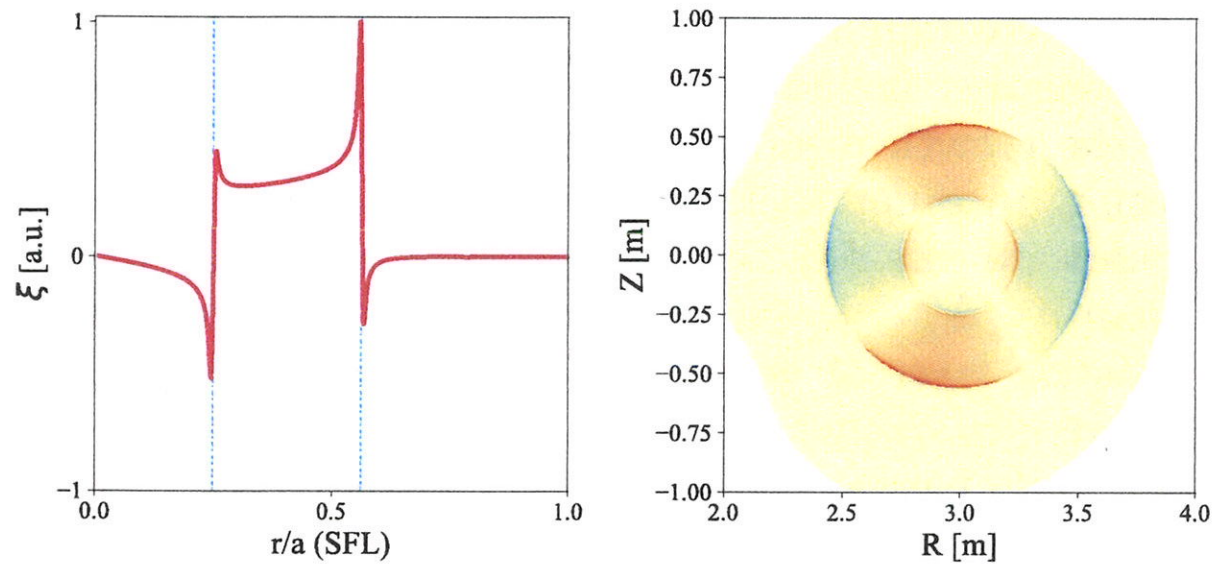

Figure 14: Radial plasma displacement structure and contour plots of the $m = n = 2$ mode in the strongly reversed shear profile: $\gamma/\omega_A = 1.3 \cdot 10^{-3}$. Parabolic pressure profile with $\alpha(r_s) = 0.02$, circular flux surfaces and $S_L = 10^6$.

When computing the tearing parameter independently of pressure, as is usually done, $D_I(r_s)$ is set to zero in (21), and the matching is straightforward after solving in the two exterior regions. However, when including finite pressure effects, the $D_I(r_s)$ term gives rise to a singularity at the rational. As a result, taking the limit $\delta_\eta \to 0$ in (23) becomes challenging, and achieving convergence is difficult.

One way to obtain convergence in $\delta_\eta$ is by using asymptotic solutions within the layer. There, $\psi$ can be expressed as two asymptotic solutions to (21), by introducing a dimensionless layer variable $x \equiv \frac{r - r_s}{r_s}$ and expanding around $x = 0$:

$$\begin{aligned} \psi_I(x) &= A_I |x|^{\nu + 1/2} - B_I |x|^{-\nu + 1/2} && \text{for } x < 0 , \\ \psi_{III}(x) &= A_{III} |x|^{\nu + 1/2} + B_{III} |x|^{-\nu + 1/2} && \text{for } x > 0 , \end{aligned} \tag{24}$$

with

$$\nu = \sqrt{\frac{1}{4} - D_I(r_s)} .$$

We used the same formalism as in Ref. [22], with the correspondence $D_s = D_I(r_s)$ and $h = \nu - 1/2$. These asymptotic solutions (24) differ from the higher order cylindrical results typically found in the literature (e.g. in [23]). This is due to the inclusion of the pressure term in $D_I(r_s)$ and the resulting singularity at $x = 0$. To compute $\Delta'_R$ we only need the ratio $A/B$ in regions $I$ and $III$, so we can fix $B_{III} = B_I = 1$. This gives:

$$\Delta'_R = A_{III} - A_I . \tag{25}$$

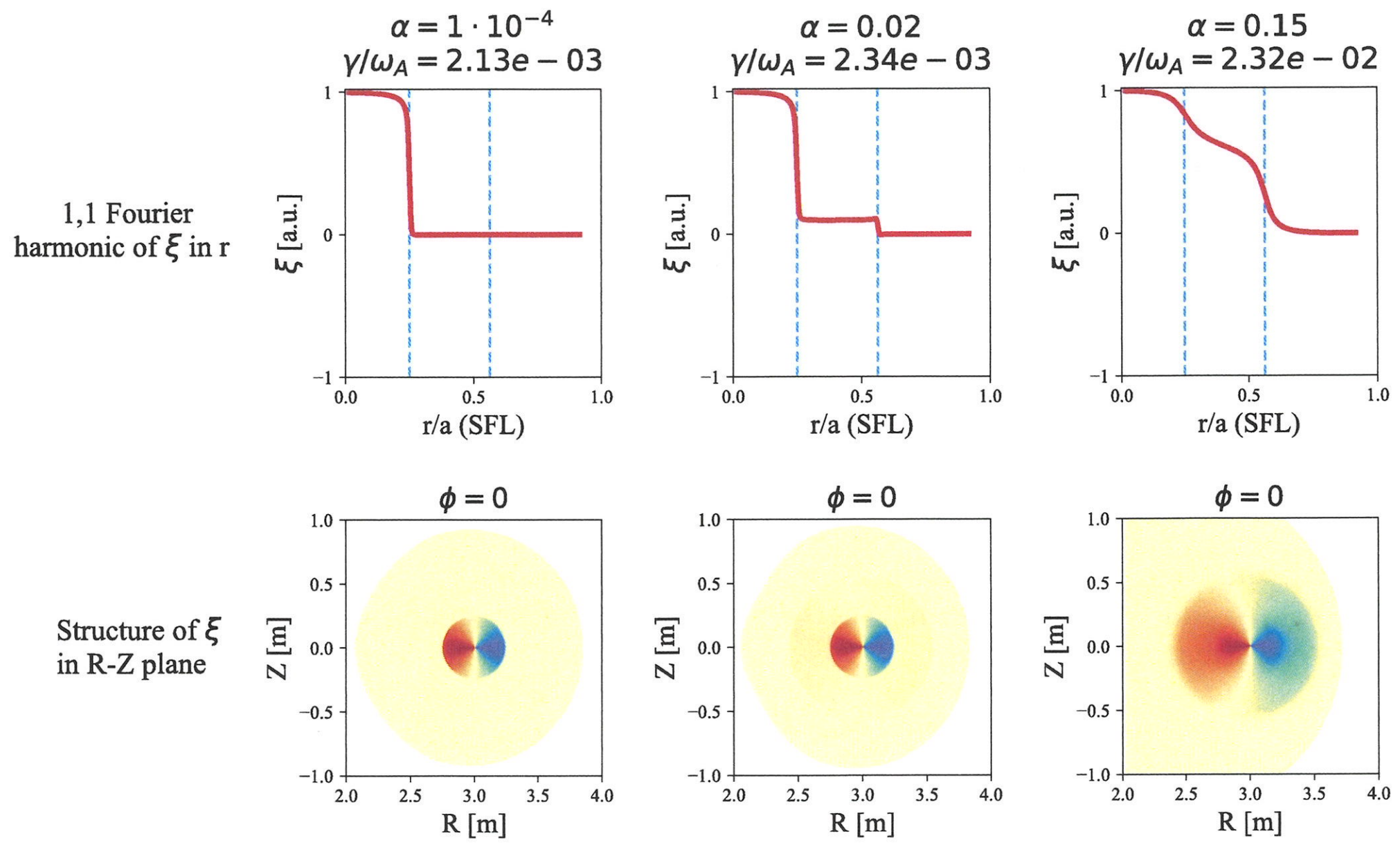

Figure 15: Radial plasma displacement structure and contour plots of the $m = n = 1$ mode for different pressure drives, in the strongly reversed shear profile. Parabolic pressure profile and circular flux surfaces; $S_L = 10^6$.

This is obtained by inserting (24) into (23) and expanding around $\nu = 1/2$ (corresponding to the zero pressure case) as well as around $\delta_\eta = 0$. We then define $\hat{\psi}'_{III}(\delta_\eta) \equiv \psi'_{III}(x = \delta_\eta)/\psi_{III}(x = \delta_\eta)$ and $\hat{\psi}'_I(-\delta_\eta) \equiv \psi'_I(x = -\delta_\eta)/\psi_I(x = -\delta_\eta)$, where the prime $'$ now denotes derivative with respect to $x$. By writing the expressions for $\hat{\psi}'_{III}(\delta_\eta)$ and $\hat{\psi}'_I(-\delta_\eta)$ using (24), we can express $A_I$ and $A_{III}$ in terms of these ratios:

$$A_I = \frac{\delta_\eta^{-1-2h}\left(-h + \delta_\eta \hat{\psi}'_I(-\delta_\eta)\right)}{1 + h + \delta_\eta \hat{\psi}'_I(-\delta_\eta)},$$
$$A_{III} = \frac{\delta_\eta^{-1-2h}\left(h + \delta_\eta \hat{\psi}'_{III}(\delta_\eta)\right)}{1 + h - \delta_\eta \hat{\psi}'_{III}(\delta_\eta)}. \tag{26}$$

The quantities $\hat{\psi}'_I(\delta_\eta)$ and $\hat{\psi}'_{III}(\delta_\eta)$ are obtained in the two exterior regions using a shooting method, with the following boundary conditions:

- Region I: shoot $\psi(r)$ following eq. (21), from $r = r_0$, with $r_0 << 1$ (we choose $r_0 = 1 \cdot 10^{-4}$) and with boundary conditions: $\psi(r_0) = r_0^m$, $\psi'(r_0) = m r_0^{m-1}$, to $r = r_s - r_s\delta_\eta$.
- Region III: shoot $\psi(r)$ following eq. (21), from $r = a$ (plasma edge), with boundary conditions: $\psi(a) = 0$, $\psi'(a) = -1$, to $r = r_s + r_s\delta_\eta$.

The shooting gives the normalized derivatives on each side of the layer: $\hat{\psi}'_{III}(x = \delta_\eta) = \psi'_{III}(r = r_s + \delta_\eta r_s)/\psi_{III}(r = r_s + \delta_\eta r_s)$ (and analogously for $\hat{\psi}'_I(x = -\delta_\eta)$). These values are then inserted into (26) to obtain $A_I$ and $A_{III}$, and finally $\Delta'_R$ using relation (25). Convergence with respect to the layer width $\delta_\eta$ should be checked.

Once $\Delta'_R$ is computed, a linear growth rate of the tearing mode can be deduced using the GGJ [18] dispersion relation:

$$\Delta'_R = \frac{2\pi\Gamma(3/4)\lambda^{5/4}}{\Gamma(1/4)L_R}\left(1 - \frac{\pi D_R(r_s)}{4\lambda^{3/2}}\right),$$
$$\text{with} \quad \lambda \equiv \gamma\tau_H^{2/3}\tau_R^{1/3} \quad \text{and} \quad L_R = r_s\tau_H^{1/3}\tau_R^{-1/3}, \tag{27}$$

where $L_R$ is the resistive interchange layer width, and the characteristic times are given by:

$$\tau_A = 1/\omega_A, \quad \tau_H = \frac{\tau_A q_s\sqrt{1 + 2q_s^2}}{sm}, \quad \text{and} \quad \tau_R = \frac{r^2}{\eta}.$$

The indicator to resistive interchange stability $D_R$, including shaping, is given by:

$$D_R(r) = \frac{\epsilon\alpha}{s(r)^2}\left[\frac{1}{q_s^2} - 1 + \frac{3}{4}(\kappa - 1)\left(1 - 2\frac{\delta}{\epsilon}\right)\right] - \frac{\alpha}{s}\Delta'. \tag{28}$$

In summary, toroidal effects can be included or neglected at different stages of the growth rate calculation: including these effects in the outer region corresponds to keeping the $D_I$ term in the ODE (21) for $\psi$. This leads to a tearing parameter depending explicitly on pressure. While this approach is more accurate, it introduces a singularity at the rational surface and makes convergence more delicate. As a result, many standard treatments (e.g. [24], [23]) choose to neglect pressure in the outer region, solving instead the simpler cylindrical equation when shooting, *i.e.* by setting $D_I(r_s) \to 0$, to obtain a constant $\Delta'_R$. More sophisticated models, such as the one proposed in [22], retain the pressure term despite the added complexity. The effect of toroidicity and shaping on $\Delta'_R$ was studied with two different codes in [25], but for pressure-free equilibria. Toroidal effects also enter the calculation in the layer (inner region), via $D_R$ in the dispersion relation (27), which accounts for finite pressure, shaping and Shafranov shift effects. These could also be neglected there, depending on the desired level of approximation. We compare the inclusion of these effects in different regions, the purpose of this section being to asses where they are most impactful, which approximations are justified, and whether it is worthwhile for codes to include them. We do so by computing the growth rates in the following four different ways:

- No pressure: 1) Obtain $\Delta'_R$ by shooting for $\psi$ in the outer region, following the ODE (21) with $D_I(r_s) \to 0$. 2) Compute the growth rate by solving the dispersion relation (27), setting $D_R(r_s) \to 0$.
- Pressure in the outer region only: 1) Obtain $\Delta'_R(\alpha)$ by shooting for $\psi$ in the outer region, following the ODE (21), including pressure by computing $D_I(r_s)$ with eq. (22). 2) Compute the growth rate by solving the dispersion relation (27), setting $D_R(r_s) \to 0$.
- Pressure in the inner region only: 1) Obtain $\Delta'_R$ by shooting for $\psi$ in the outer region, following the ODE (21) with $D_I(r_s) \to 0$. 2) Compute the growth rate by solving the dispersion relation (27), including pressure by computing $D_R(r_s)$ with eq. (28).
- Pressure everywhere: 1) Obtain $\Delta'_R(\alpha)$ by shooting for $\psi$ in the outer region, following the ODE (21), including pressure by computing $D_I(r_s)$ with eq. (22). 2) Compute the growth rate by solving the dispersion relation (27), including pressure by computing $D_R(r_s)$ with eq. (28).

We compare the inclusion/exclusion of these effects in the inner and outer region in Figure 16, for a $m/n = 2/1$ tearing mode. The figure also shows growth rates from the resistive modular solver, labelled as "Full equations solution". As a reminder, in this solver, toroidal effects are included throughout and there is no need for matching between the inner and outer regions. The calculations are made for two different types of pressure profiles (shown along the $q$ profile in Fig. 17). The first one is a broad parabolic pressure profile, and the second is a sharper profile, with the maximum pressure gradient located at the rational surface. We also explore shaping effects, with Figure 16 displaying a case with slight negative triangularity on the right.

For these different cases, the inclusion of pressure in the outer region does not appear to significantly impact the growth rate. This is consistent with the findings of [22], which used a parabolic pressure profile and circular flux surfaces, and investigated the effect of including pressure in the outer region. They recommand considering a constant tearing parameter unless in a very high $\beta$ regime. We observe that this approximation holds for sharper pressure profiles and shaped flux surfaces, with virtually no impact on the growth rate, apart from a very slight artificial destabilisation at high $\beta$. However it is clear that including pressure in the inner region when calculating GGJ growth rates is crucial to obtaining the correct trend with $\alpha$, regardless of the pressure profile or shaping. Neglecting pressure in the inner region leads to loss of the pressure stabilisation effect on tearing modes. To conclude, while it is essential to include pressure in the inner region (in the GGJ dispersion relation (27)), using a constant value for the tearing parameter $\Delta'_R$, computed with zero pressure, not only saves many numerical difficulties but is also a reasonable approximation for obtaining consistent growth rates.

### B. Impact of inconsistent inclusion of $\delta B_\parallel$ and equilibrium toroidal field

Reference [16] quantified the impact of not including parallel magnetic field fluctuations $\delta B_\parallel$ consistently for all pressure driven ideal modes in tokamaks. We use a few equations of [16], starting with the definition of $\delta B_\parallel$ in terms of the radial plasma displacement, at leading order in $\epsilon$:

$$\delta B_\parallel = -\frac{B_0}{R}\xi^r\left(\frac{\alpha}{2q^2} + \epsilon\left[\left(\frac{n}{m} + \frac{1}{q}\right)\left(\frac{n}{m} - \frac{1}{q}\right)\right]\right).$$

Some gyrokinetic and nonlinear reduced MHD codes, such as EUTERPE , ORB5 [26], GTC [27] or TM1 [28] adopt the following assumption about the perturbed vector potential: $\boldsymbol{\delta A} = \delta A_\parallel \boldsymbol{b}$ . Under this constraint, the resulting $\delta B_\parallel$ in these partially electromagnetic codes is almost zero, given by :

$$\widetilde{\delta B_\parallel} = \frac{B_0}{Rq^2}\epsilon\xi^r\left(1 - \frac{nq}{m}\right)(2 - s)\,.$$

This substantial diminution in parallel magnetic field fluctuations modifies the Mercier criterion as follows:

$$s^2\widetilde{D_I} = \epsilon\alpha\left[\frac{1}{q_s^2} - 1 + \frac{3}{4}(\kappa - 1)\left(1 - 2\frac{\delta}{\epsilon}\right)\right] - q_s^2\left[\frac{\alpha}{2q_s^2} + \epsilon\left(\left(\frac{1}{q_s} - \frac{n}{m}\right)^2 - \frac{s}{q}\left(\frac{1}{q_s} - \frac{n}{m}\right)\right)\right]^2 .$$

The entire second term in $-q_s^2$ is an additional contribution absent from the correct Mercier term $D_I$ (given in eq. (1) in section II). This term arises from the inconsistent treatment of $\delta B_\parallel$, and would otherwise cancel out. In this extra term, the $\frac{\alpha}{2q_s^2}$ component dominates, while the $\epsilon$ term is of order $O(\epsilon^2\Delta q^2)$. As argued in [16], we discard this second term, which is analogous to field line bending, but is higher order, and thus does not compete in the global governing equations. We therefore consider the following modified Mercier term:

$$s^2\widetilde{D_I} = \epsilon\alpha\left[\frac{1}{q_s^2} - 1 + \frac{3}{4}(\kappa - 1)\left(1 - 2\frac{\delta}{\epsilon}\right)\right] - \frac{\alpha^2}{4q_s^2}\,. \quad (29)$$

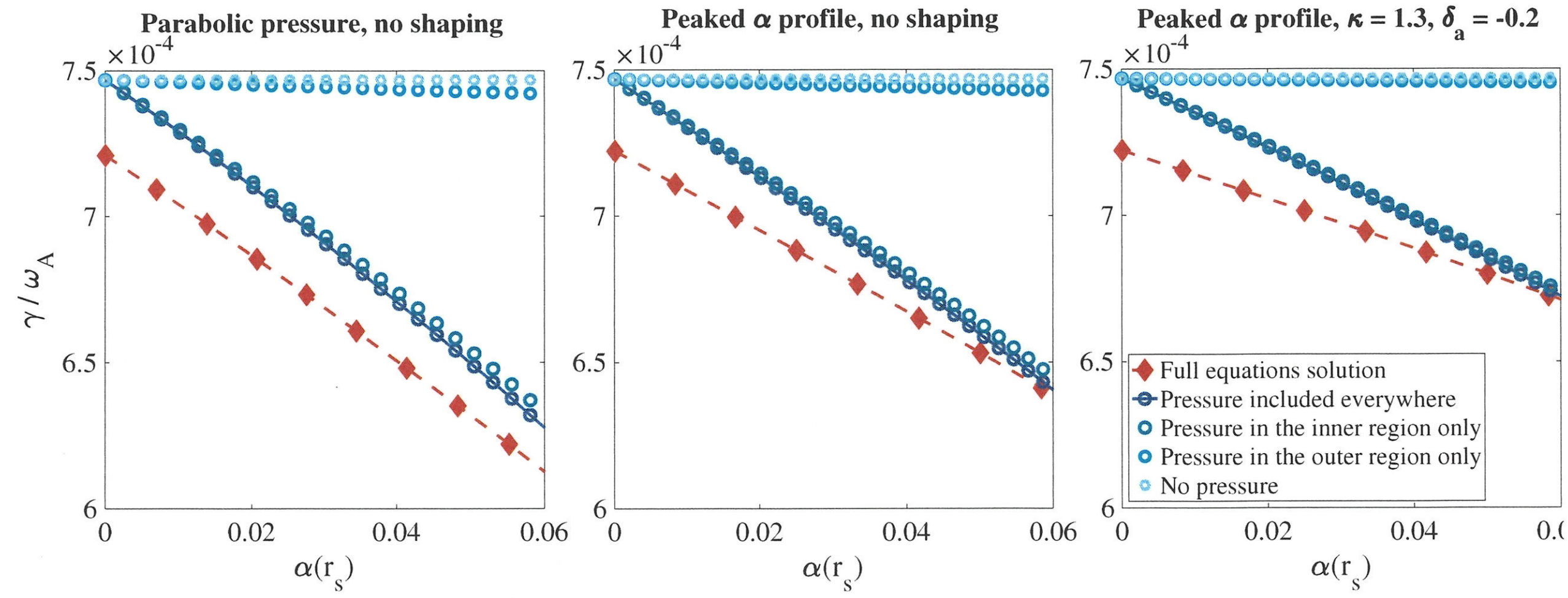

Figure 16: Growth rates of a $m/n = 2/1$ tearing mode, comparison of modular solver (full equations) results, and different ways of solving the GGJ dispersion relation. The comparison is performed for different types of pressure profiles, and shaping of the flux surfaces (profiles given in Fig. 17).

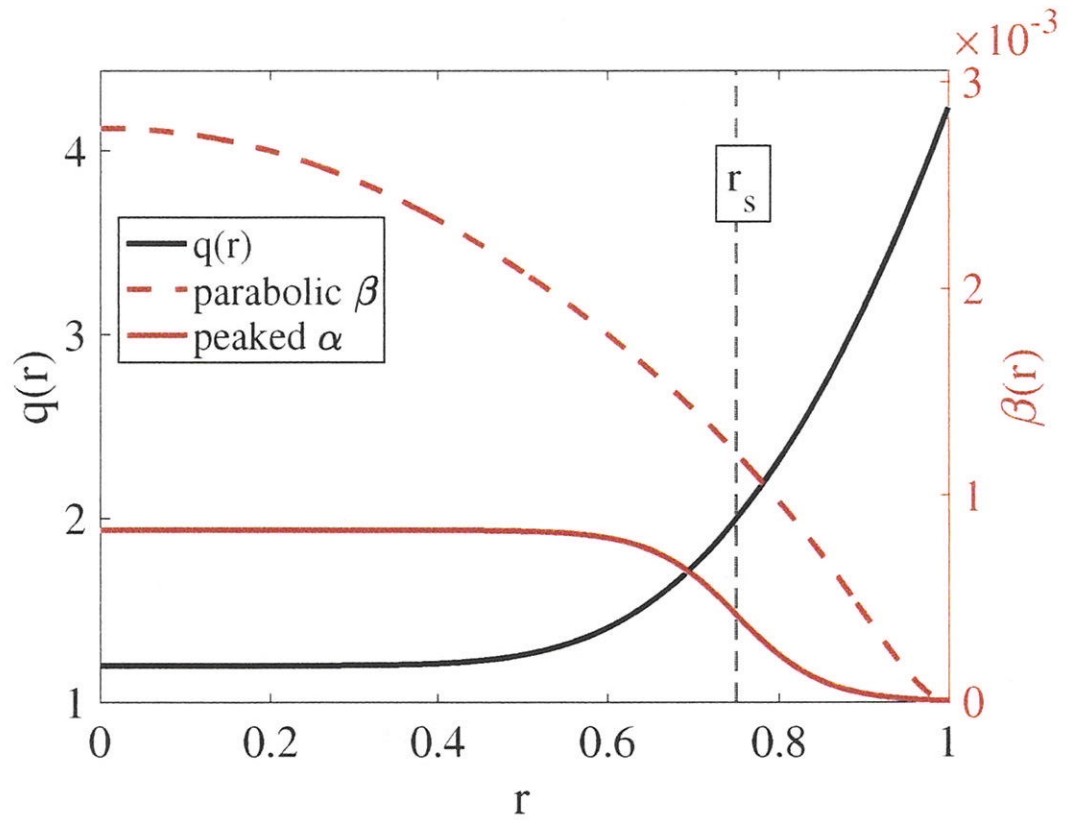

Figure 17: Pressure and $q$ profiles used to obtain the $m/n = 2/1$ tearing mode growth rates given in Fig. 16. The $\beta$ profiles are given for $\alpha(r_s) = 0.05$.

*1. Effect of setting $\delta A_\perp \to 0$ on interchange/infernal growth rates*

The impact of this approximation on ideal and resistive interchange can be seen immediately by recomputing the Mercier criterion with the additional terms. The analytical criterion for ideal interchange instability is $D_I > 1/4$, giving:

$$\begin{aligned}
&\frac{\epsilon\alpha}{s^2}\left[\frac{1}{q_s^2} - 1 + \frac{3}{4}(\kappa - 1)\left(1 - 2\frac{\delta}{\epsilon}\right)\right] > \frac{1}{4} : \quad \text{ideal interchange unstable for consistent } \delta B_\parallel\,, \\
&\frac{\epsilon\alpha}{s^2}\left[\frac{1}{q_s^2} - 1 + \frac{3}{4}(\kappa - 1)\left(1 - 2\frac{\delta}{\epsilon}\right)\right] > \frac{1}{4} + \frac{\alpha^2}{4s^2q_s^2} : \quad \text{ideal interchange unstable for } \delta A_\perp \to 0\,.
\end{aligned} \quad (30)$$

For resistive interchange, the usual criterion $D_R \equiv D_I - \alpha\Delta'/s > 0$ translates to:

$$
\begin{aligned}
&\frac{\epsilon\alpha}{s^2}\left[\frac{1}{q_s^2} - 1 + \frac{3}{4}(\kappa - 1)\left(1 - 2\frac{\delta}{\epsilon}\right)\right] - \frac{\alpha\Delta'}{s} > 0 \ : \quad \text{resistive interchange unstable for consistent } \delta B_\parallel\,, \\
&\frac{\epsilon\alpha}{s^2}\left[\frac{1}{q_s^2} - 1 + \frac{3}{4}(\kappa - 1)\left(1 - 2\frac{\delta}{\epsilon}\right)\right] - \frac{\alpha\Delta'}{s} > \frac{\alpha^2}{4s^2q_s^2} \ : \quad \text{resistive interchange unstable for } \delta A_\perp \to 0\,.
\end{aligned}
\tag{31}
$$

It is straightforward to see that it will be much harder to be interchange unstable when $\delta A_\perp = 0$. For resistive interchange, which is normally always unstable for $q_s < 1$, even at zero pressure, there will be an artificial stabilization over a wide range of parameters. These analytic criteria (30) and (31) are represented by the red lines in Figs. 18 and 19, for a better representation of the change in marginal stability across a large range of $\alpha$ and magnetic shear.

The situation for infernal modes is more intricate, as higher-order terms must be retained in the equations, and we cannot easily use a simple analytic criterion, especially for resistive cases. Using the modular solver to solve the full equations allows to quantify the effect of neglecting $\delta B_\parallel$ on infernal modes: in a modified version of the solver, the Mercier term is modified accordingly in the full equations to simulate the approximation $\delta A_\perp \to 0$. Results for a $m/n = 9/10$ case are given in Fig. 18 for the ideal problem, and Fig. 19 for the resistive one. The dashed white line represents the marginal stability found with the code (set at $\gamma/\omega_A = 5 \cdot 10^{-5}$). The stability situation is clearly completely different when $\delta B_\parallel$ is included consistently or not. In the reduced model $\delta A_\perp \to 0$, infernal modes are somewhat easier to observe than interchange modes. This is seen for the ideal case in Fig. 18b, and for the resistive case Fig. 19b, comparing the adapted analytic interchange criterion (marginal point) shown in red to the infernal growth rate which appear in colour. Thus, the feature of infernal modes being easier to destabilize than typical interchange modes is preserved in the partially electromagnetic model. However the resulting growth rates are much smaller (the scale is the same on both sides of Figs. 18 and 19). Moreover, as magnetic shear increases, ideal infernal modes are completely artificially stabilized in the reduced model case. For resistive infernal modes, for small values of $\alpha$ it is still possible to observe them as magnetic shear increases (19b). The marginal stability obtained with the code (in white) recovers the analytic criterion (in red): for small $\alpha$ ($\alpha \lessapprox 0.05$), assuming $\delta A_\perp = 0$ is not so harmful and resistive interchange/infernal modes can still be observed even for higher values of magnetic shear. However, increasing $\alpha$, we quickly reach marginal stability, even though resistive infernal modes should always remain unstable for this $m/n = 9/10$ case, as seen in 19a.

In summary, it is very difficult, nearly impossible, to observe resistive or ideal interchange or infernal modes with the reduced model $\delta A_\perp \to 0$, even though solving the full equations indicates that these modes should be dominant with very large growth rates. Treating the resistive problem with $\delta A_\perp = 0$ may be consistent for small values of $\alpha$, but in general, the two models only converge for very small values of both $\alpha$ and magnetic shear, which is expected given the modifications to the Mercier term given in equation (29).

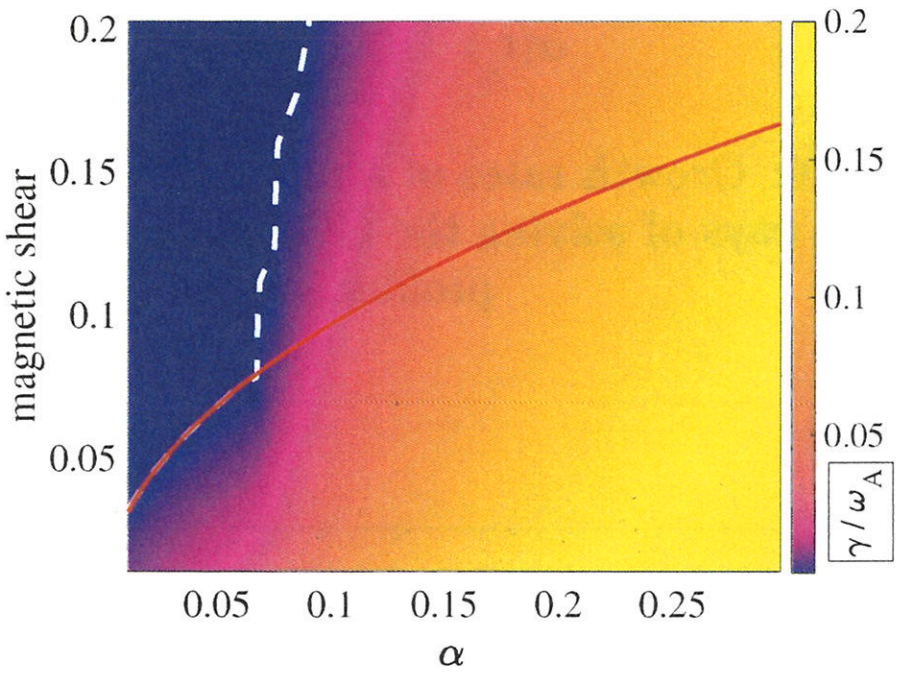

(a) Solving consistently for $\delta B_\parallel$

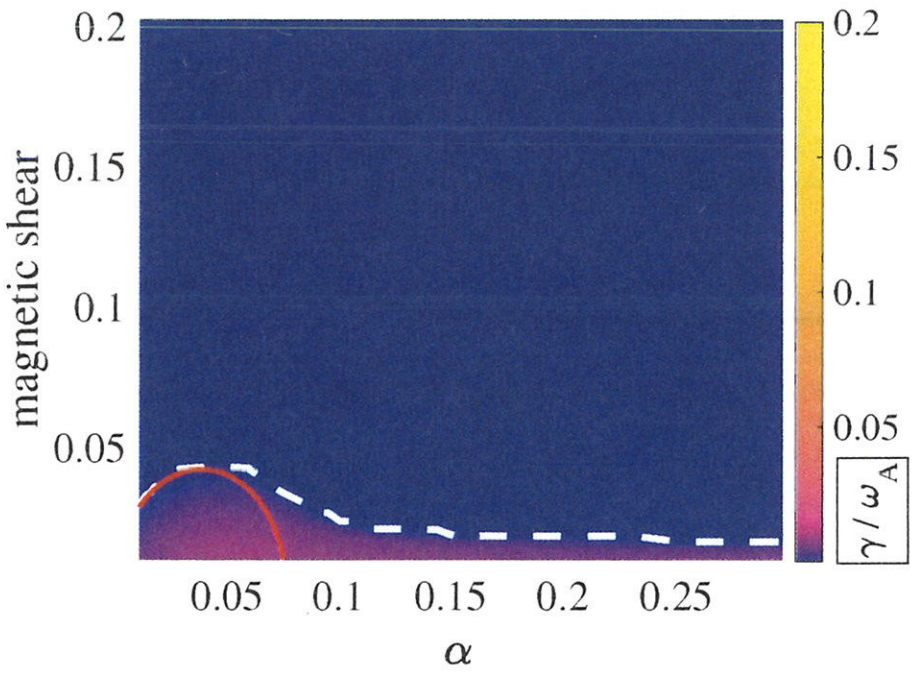

(b) $\delta A_\perp \to 0$

Figure 18: Influence of $\delta B_\parallel$ corrections on $m/n = 9/10$ ideal growth rates obtained with the modular solver. White line is marginal stability ($\gamma/\omega_A = 5 \cdot 10^{-5}$). Red line is Mercier criterion adapted (30). The pressure and $q$ profiles used are given in Appendix (D1) and (D2) respectively.

*2. Effect of setting $\delta A_\perp \to 0$ on tearing mode growth rates*

What about tearing modes observation in this reduced model? Once again, we use the modular solver to obtain growth rates and investigate the impact of the additional

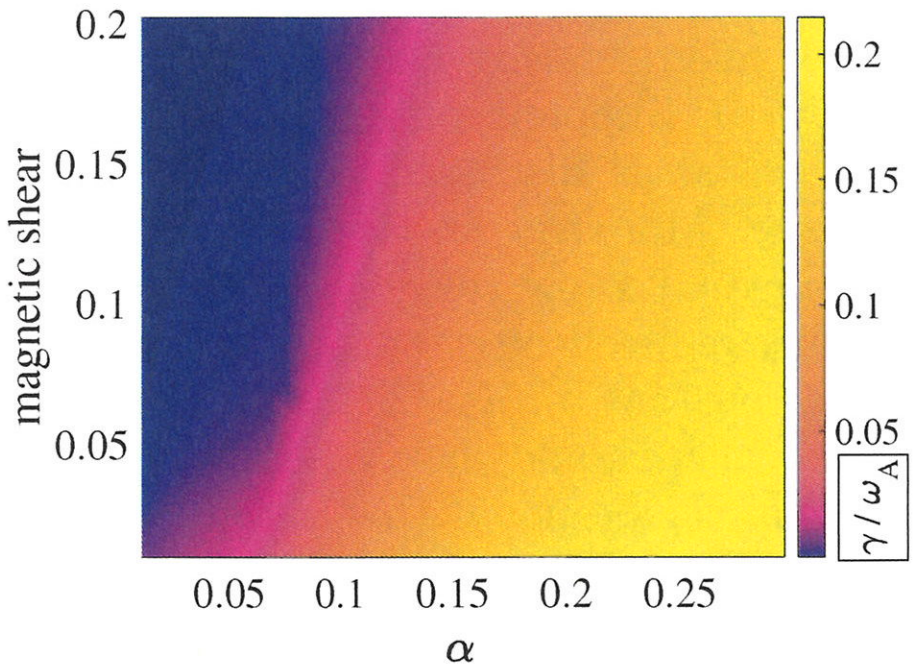

(a) Solving consistently for $\delta B_\parallel$

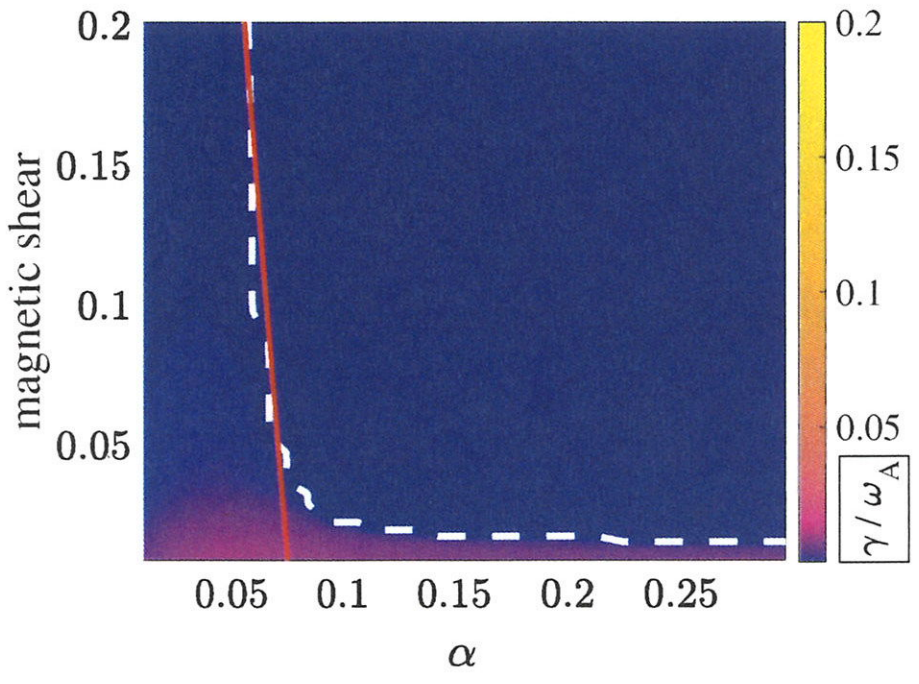

(b) $\delta A_\perp \to 0$

Figure 19: Influence of $\delta B_\parallel$ corrections on $m/n = 9/10$ resistive ($S_L = 10^7$) growth rates obtained with the modular solver. White line is marginal stability ($\gamma/\omega_A = 5 \cdot 10^{-5}$). Red line is Mercier criterion adapted (31). The pressure and $q$ profiles used are given in Appendix (D1) and (D2) respectively.

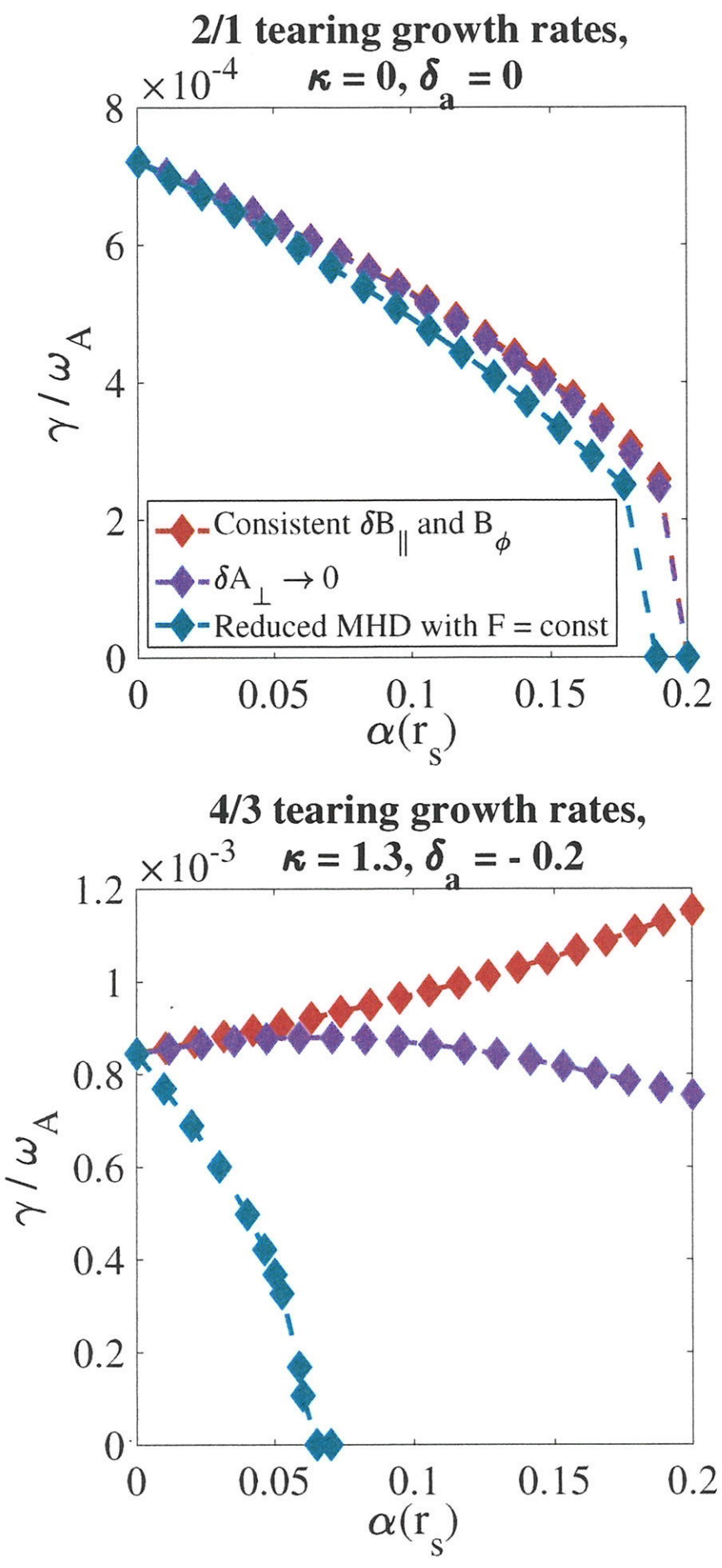

Figure 20: Tearing mode growth rates for the $q$ profile shown in black in Fig. 7 (same for both panels), parabolic pressure profile and $S_L = 10^6$.

$\alpha^2/4q_s^2$ term. Figure 20 shows tearing mode growth rates, comparing cases where $\delta B_\parallel$ solved consistently (in red) to cases assuming $\delta A_\perp = 0$ (in purple). For the typical $m/n = 2/1$ tearing with circular flux surfaces shown on top in Fig. 20, we can see that the assumption $\delta A_\perp = 0$ has almost no impact on the growth rate. The marginal point in $\alpha$ remains unchanged, and while there is a slight artificial stabilisation in the reduced model, it is completely negligible. The situation is completely different from the interchange and infernal modes studied previously. While those modes are pressure-driven, tearing modes are typically stabilised by pressure gradients. As a result, increasing $\alpha$ leads to a reduction in the growth rate, eventually reaching marginal stability before the effect of the $\alpha^2/4q_s^2$ term becomes significant.

If we look at a smaller $q_s$, however, the extra term becomes more important and may have a larger impact on the growth rates. Changing the shaping of the flux surfaces also modifies the growth rate dependence on $\alpha$. Sufficient negative triangularity changes the sign of the magnetic well, and pressure gradients can become destabilizing rather than stabilizing for tearing modes (this can be seen in Figure 6 for example). This is observed in the bottom panel of Figure 20, for a $m/n = 4/3$ tearing with $\kappa = 1.3$ and $\delta_a = -0.2$. When $\delta B_\parallel$ is solved consistently (shown in red), the growth rate increases with $\alpha$. For this type of scenario, assuming $\delta A_\perp = 0$ has now an important impact. As seen on the purple curve, this assumption artificially stabilizes the mode with pressure, leading to a completely different trend from the correct one, where the growth rate should increase with pressure. There is only a narrow $\alpha$ window, around $\alpha(r_s) \sim 1 \cdot 10^{-3}$, where the assumption has negligible effect and the growth rates remain similar.

### 3. *Impact of assuming constant equilibrium toroidal field on tearing modes growth rates*

Some nonlinear MHD codes [29], in addition to simplify electromagnetic perturbations by setting $\delta A_\perp$ as just discussed, also approximate the equilibrium magnetic field in the stability equations. We can write the equilibrium field as:

$$\boldsymbol{B} = F(\psi)\nabla\phi + \nabla\phi \times \nabla\psi ,$$

where $F(\psi)$ is normally a flux function, but in this reduced model it is assumed constant: $F = R_0 B_0$. Reference [16] examined the combined impact of these two assumptions (see Section 5), on ideal pressure-driven modes, notably concluding that interchange modes would only be found unstable for negative $\alpha$. The change in the Mercier term in this approximation is found to be:

$$s^2\widetilde{\widetilde{D_I}} = \epsilon\alpha\left[-1 + \frac{3}{4}(\kappa - 1)\left(1 - 2\frac{\delta}{\epsilon}\right)\right] - \frac{\alpha^2}{4q_s^2} . \quad (32)$$

Running a modified version of the solver incorporating the updated Mercier term (32), we simulate both $\delta A_\perp \to 0$ and $F = R_0 B_0$. Results for the 2/1 and 4/3 tearing modes are shown in blue in Figure 20, alongside consistent growth rates (in red) and those assuming only $\delta A_\perp \to 0$ (in purple). The effect of this reduced model follows the same trend as assuming only $\delta A_\perp \to 0$: an artificial stabilisation of tearing mode growth rates with pressure. This effect becomes dramatic with negative triangularity shaping of the flux surfaces (bottom panel of Fig. 20), where the tearing mode is completely stabilised at a low value of $\alpha$, while the consistent growth rate is not only significant, but continues to increase with $\alpha$. These results show that combining a constant toroidal field with the reduced electromagnetic perturbation leads to a much more severe deviation than assuming $\delta A_\perp$ alone.

In summary, assuming $\delta A_\perp = 0$ essentially prevents the observation of pressure driven modes such as interchange or infernal modes altogether, even where they should be dominant with large growth rates, unless in a very narrow low-$\alpha$ regime. As found analytically in Ref. [16] for the ideal case, a consistent treatment of $\delta B_\parallel$ is therefore crucial to correctly capturing these types of instabilities. In the current work we also examine tearing modes: for circular flux surfaces, or in the presence of positive triangularity (which further stabilises the mode with pressure), and if $q_s = m/n$ is relatively high (as in the $m/n = 2/1$ case), assuming $\delta A_\perp = 0$ is not problematic and does not significantly affect the growth rate. Even when combined with the assumption of a constant toroidal field, the discrepancy remains limited at moderate $\alpha$. However, with negative triangularity and / or lower $q_s$, the $\delta A_\perp = 0$ simplification leads to incorrect results as $\alpha$ increases. In such cases, it is essential to compute $\delta B_\parallel$ consistently. This result could be very important for gyrokinetic calculations of micro tearing modes. Such codes typically adopt $\delta A_\perp \to 0$. Simple replacement of grad-B drifts by curvature drifts, as discussed in Ref. [16] should correct the issue. Moreover, in these negative triangularity scenarios, beyond a low treshold in pressure, codes that combine this reduced electromagnetic formulation with assuming a constant toroidal field will entirely miss tearing modes which should have a significant growth rate. Finally, it should be noted that the impact of $\delta A_\perp \to 0$ and $F = R_0 B_0$ is important only for modes where $q_s$ is not much larger than unity. Mercier corrections are not very important for example for peeling-ballooning modes. These modes are the main interest for reduced MHD codes.

## V. CONCLUSIONS

This paper explores the stability properties of advanced tokamak regimes, which feature extended regions of low magnetic shear. These configurations, though promising for future reactors, are more susceptible to the development of pressure-driven instabilities known as (ideal or resistive) infernal modes. In some cases, a full spectrum of infernal modes can arise for a single poloidal and toroidal mode number $(m, n)$. This paper aims to systematically investigate the influence of different parameters (resistivity, flux surface shaping, and toroidal effects) on the stability of such scenarios and spectra, with a particular focus on reversed shear $q$ profiles.

To this end, we rely on both analytic dispersion relations and a modular resistive MHD eigenvalue solver [2] based on a unified and global description of pressure and current driven internal instabilities [1]. This model is now extended to include shaping effects through modifications of the Mercier well, as described in the first section. We first studied non-resonant $q$ profiles, for which ideal MHD is a good description. The large variety of modes and different spectra observed with the modular solver motivated us to derive analytic dispersion relations, to be able to predict whether a spectrum appears or not, and calculate the corresponding growth rates. A dispersion relation had already been derived for monotonic non-resonant $q$ profiles in [2], with circular flux surfaces. The first step was to extend this dispersion relation to include shaping. The resulting analytic description captures the key features of the infernal spectrum without requiring the use of the modular solver. In particular, shaping was found to profoundly affect the stability landscape: even moderate modifications of elongation or triangularity could either suppress the instability spectrum entirely or trigger the appearance of an infinite number of unstable modes. To be able to study reversed $q$ profiles, we then extended the description to sheared profiles. Using an energy minimization approach, we derived a relatively simple dispersion relation to predict the emergence of spectra and associated growth rates in shaped, reversed $q$ profiles, showing good agreement with the modular solver. Unstable interchange "Mercier-like" criteria were also obtained. These expressions offer a practical way to assess the stability of weak and reversed shear configurations. We find that in low shear regions, these profiles give rise to the same kind of spectrum as for monotonic $q$ profiles, with slightly more complex plasma displacements. The eigenfunctions, having infernal mode structure enable the derivation of an alternative treatment of interchange modes to that traditionally adopted (*e.g.* by Mercier), not restricted to the most unstable mode but allowing the description of the full spectrum.

The second part of the work focused on resistive phenomena. We investigated first monotonic profiles featuring a $q_s = 1$ rational surface in the core, relevant for advanced inductive scenarios, or for certain hybrid scenarios. We find that the combined effect of negative triangularity (even a small amount) and resistivity can give rise to spectrum of instabilities that would otherwise not be there. Adding on a second $q_s = 1$ rational surface, we study the difference between a weakly and a strongly reversed profiles, for different poloidal mode numbers $m$. At low pressure, the strong reversal case is more unstable across all mode numbers, except at high $m$ and negative triangularity. As the pressure increases, however, the weak shear scenario becomes more unstable overall. The weakly sheared configuration reveals a transition in the eigenmode structure, going from a typical tearing mode shape to a resistive infernal mode one as pressure increases.

This variety of modes arising in low-shear configurations can sometimes be lost due to assumptions in numerical codes. In the last section, we examine how common numerical simplifications can diminish or suppress these instabilities. We first investigated different approaches for calculating growth rates using the GGJ dispersion relation and the tearing parameter $\Delta'_R$. We examine the impact of including toroidal effects (such as pressure, shaping, and Shafranov shift) in the inner (resistive layer) or outer region in the calculation. Our results show that including pressure effects in the inner region is crucial to retrieve correct growth rate, whereas the tearing parameter can be computed with zero pressure without significant loss of accuracy. We also assessed partially electromagnetic models which assume $\boldsymbol{\delta A} = \delta A_\parallel \boldsymbol{b}$ for the perturbed vector potential. While adequate for tearing modes in circular or positively triangular plasmas, this approximation severely underestimates growth rates in low $q_s$ or negative triangularity configurations. This is even worse for reduced models which also assume a constant toroidal equilibrium field: even at low pressure, fast-growing tearing modes are incorrectly considered stable. Setting $\boldsymbol{\delta A} = \delta A_\parallel \boldsymbol{b}$ also completely prevents the observation of ideal and resistive pressure-driven modes such as interchange or infernal modes, except in extremely low-$\alpha$ regimes. Proper computation of $\delta B_\parallel$ thus proved essential for correctly capturing the physics of these instabilities, except for very low pressure gradients simulations.

Overall, advanced low-shear scenarios, while attractive for next-generation devices, present a broad range of physical phenomena and instabilities. With two rational surfaces, the eigenmode structure in weakly reversed shear scenario shows an interesting transition between double tearing modes and resistive infernal modes. With negative triangularity, in addition to the loss of tearing mode stabilisation by pressure, the core is much more MHD unstable and prone to exotic physics. The combination of negative triangularity and low magnetic shear give rise to spectra of ideal or resistive infernal modes in regions of vanishing or favourable average curvature. The solutions behaviour is completely different with or without shaping. Through this detailed investigation, we hope to contribute to the theoretical framework necessary for understanding and controlling the stability of advanced tokamak regimes, and to provide a foundation for future studies and the development of robust scenarios. The spectra uncovered in this study are not necessarily detrimental for overall stability, as they could potentially play a beneficial role by leading to reconnection in the core, and avoiding long wavelength instabilities such as sawteeth.

## Acknowledgements

This work has been carried out within the framework of the EUROfusion Consortium, via the Euratom Research and Training Programme (Grant Agreement No 101052200 — EUROfusion) and funded by the Swiss State Secretariat for Education, Research and Innovation (SERI). Views and opinions expressed are however those of the author(s) only and do not necessarily reflect those of the European Union, the European Commission, or SERI. Neither the European Union nor the European Commission nor SERI can be held responsible for them.

## Appendix A: Constant shaping coefficients

The penetration of shaping into the core is governed by the following equation (to leading order in aspect ratio):

$$r^2 S''_m(r) + (3 - 2s(r)) r S'_m(r) + (1 - m^2) S_m(r) = 0\,, \tag{A1}$$

where the shaping coefficients used in this work are defined as:

$$\kappa(r) = \frac{r - S_2(r)}{r + S_2(r)},$$

$$\delta(r) = \frac{4}{r} S_3(r).$$

For very low shear cases, one can take $s(r) \to 0$ in equation (A1), and obtain:

$$\kappa(r) = \kappa(a), \tag{A2}$$

$$\delta(r) = \delta(a)\frac{r}{a}. \tag{A3}$$

This approximation is used throughout this work, as we focus on low-shear configurations. In Section II, where $q$ profiles are nearly flat for $r < r_1$, we use $\kappa = \kappa(r_1)$ and $\delta(r) = \delta_1 r/r_1$, making the approximation nearly exact.

However, we briefly consider two tearing mode growth rate calculations (involving higher shear). One of these cases is shown in Fig. 20, where the $q$ profile used was:

$$q(r) = q_0\left(1 + cr^{3d}\right)^{1/d}, \tag{A4}$$

with $q_0 = 1.2$, $\tilde{r} = 0.75$, $\tilde{q} = 2$, $d = 2.5$, and $c$ given by:

$$c = \left(\left(\frac{\tilde{q}}{q_0}\right)^d - 1\right)\tilde{r}^{-3d}.$$

For a $q$ profile of the type (A4), solving (A1), and keeping the two first order terms when expanding around $r = 0$ gives:

$$\kappa(r) = \frac{d(4+3d)\kappa_a + c\left(r^{3d}(\kappa_a - 1) + 1 + \kappa_a\right)}{d(4+3d) + c\left(r^{3d}(\kappa_a - 1) + 1 - \kappa_a\right)},$$

$$\delta(r) = r\,\delta_a\frac{3d(2+d) + 4cr^{3d}}{3d(2+d) + 4c}.$$

Figure 21 shows the shear profile for the case of Fig. 20, along with the corresponding shaping penetration into the core. In tearing mode scenarios, the shear is significantly higher than in the rest of the cases studied in this paper. As a result, the comparison between the actual solutions for $\kappa(r)$ and $\delta(r)$ and the approximations $\kappa(r) = \kappa_a$ and $\delta(r) = \delta_a r/a$ becomes less accurate. Nevertheless, this is a moderate discrepancy. Furthermore, as discussed in section IV A, the impact of toroidal effects (including shaping) on stability is primarily localized near the rational surface, and negligible in the outer region. This supports the validity of our approximation, as shaping has rather a local influence.

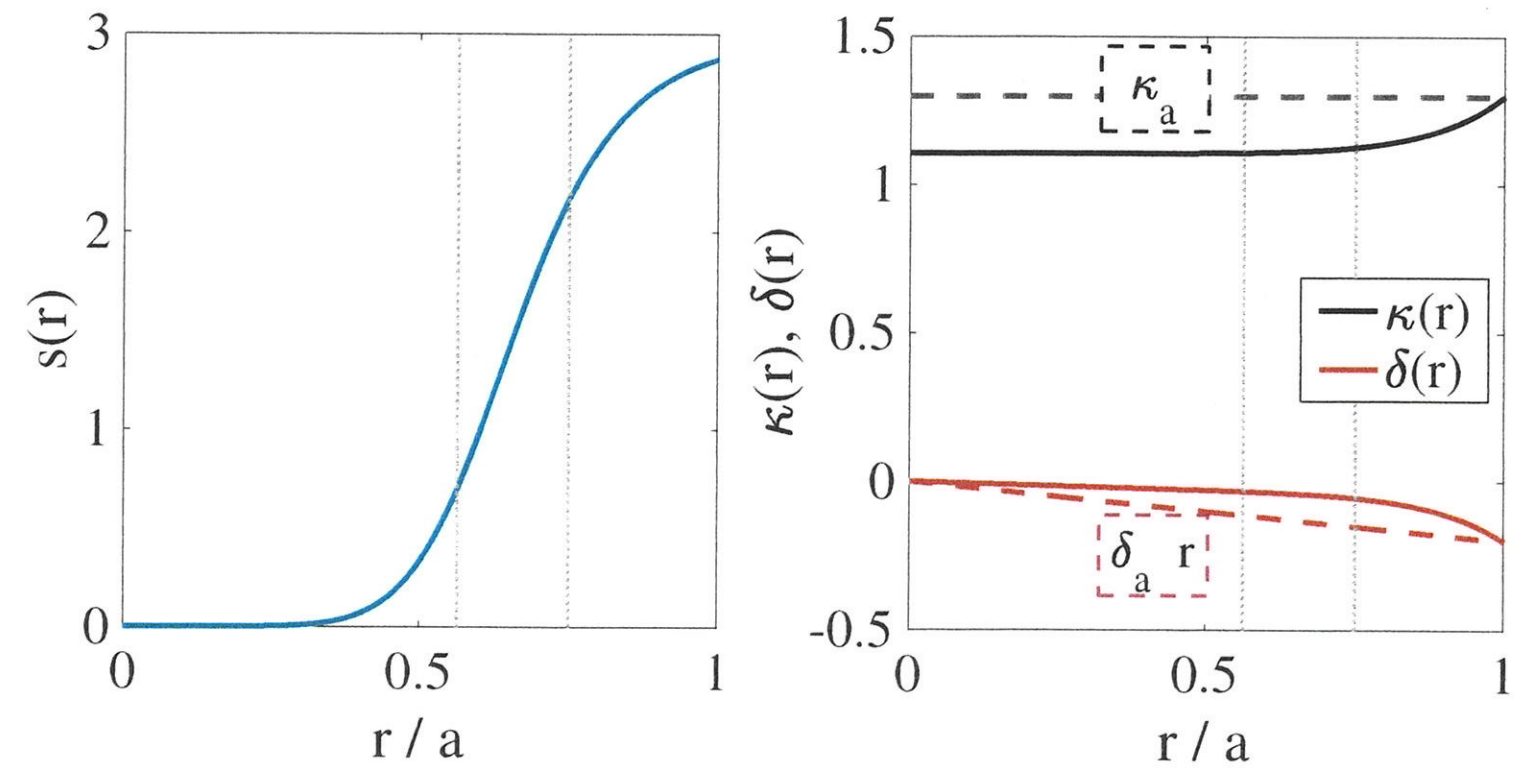

Figure 21: Shaping penetration with into the core: approximations (A3) (dashed lines) vs. solutions of eq. (A1) (solid lines), for the $q$ profile of Fig. 20. $\kappa_a = 1.3$ and $\delta_a = -0.2$. Dashed lines indicate rational surfaces where $q = 4/3$ and $q = 2/1$.

## Appendix B: Equations behind the modular solver

The modular solver we use in this work is based on a linearized MHD description of internal instabilities in toroidal geometry, accounting for resistive diffusion, compressibility, and basic shaping effects. A static axisymmetric equilibrium is perturbed. Poloidal components are eliminated to solve for the radial plasma displacement $\xi$, defined as $\delta v = \delta\xi/\delta t$, where $v$ is the radial velocity. The equations were first derived in [1], and their adaptation in a solver (as well as benchmarking) is described in [2]. The problem is solved as a generalized eigenvalue problem, with eigenvector: $(\xi^{(m)}, \gamma/\omega_A\xi^{(m)}, \chi^{(m)}, \Delta\xi_\Gamma, \xi^{(m+1)}, \gamma/\omega_A\xi^{(m+1)}, \chi^{(m+1)}, \xi^{(m-1)}, \gamma/\omega_A\xi^{(m-1)}, \chi^{(m-1)})$, which can be represented as in (B1):

$$
\begin{aligned}
\frac{\gamma}{\omega_A}\mathcal{D}\left(\frac{\gamma}{\omega_A}\xi^{(m)}\right) &= \mathcal{D}_\xi(\xi^{(m)}) + \mathcal{D}_+(\bar{\xi}^{(m+1)}) + \mathcal{D}_-(\bar{\xi}^{(m-1)}) \\
&\quad + \mathcal{D}_\chi(\chi) + \mathcal{D}_{\Delta\xi}(\Delta\xi_\Gamma) \\
\frac{\gamma}{\omega_A}\mathcal{O}(\chi) &= \mathcal{O}_\xi(\xi^{(m)}) + \mathcal{O}_\chi(\chi) \\
\frac{\gamma}{\omega_A}\mathcal{S}^+\left(\frac{\gamma}{\omega_A}\bar{\xi}^{(m+1)}\right) &= \mathcal{S}^+_\xi(\xi^{(m)}) + \mathcal{S}^+_+(\bar{\xi}^{(m+1)}) + \mathcal{S}^+_\chi(\chi) \\
&\quad + \mathcal{S}^+_{\Delta\xi}(\Delta\xi_\Gamma) \\
\frac{\gamma}{\omega_A}\mathcal{O}^+(\chi^{m+1}) &= \mathcal{O}^+_+(\bar{\xi}^{(m+1)}) + \mathcal{O}^+_{\chi^+}(\chi^+) \\
\frac{\gamma}{\omega_A}\mathcal{S}^-\left(\frac{\gamma}{\omega_A}\bar{\xi}^{(m-1)}\right) &= \mathcal{S}^-_\xi(\xi^{(m)}) + \mathcal{S}^-_-(\bar{\xi}^{(m-1)}) + \mathcal{S}^-_\chi(\chi) \\
&\quad + \mathcal{S}^-_{\Delta\xi}(\Delta\xi_\Gamma) \\
\frac{\gamma}{\omega_A}\mathcal{O}^-(\chi^{m-1}) &= \mathcal{O}^-_-(\bar{\xi}^{(m-1)}) + \mathcal{O}^-_{\chi^-}(\chi^-) \\
\frac{\gamma}{\omega_A}\mathcal{C}\left(\frac{\gamma}{\omega_A}\Delta\xi_\Gamma\right) &= \mathcal{C}_\chi(\chi) + \mathcal{C}_{\Delta\xi}(\Delta\xi_\Gamma)\,,
\end{aligned}
\tag{B1}
$$

where we changed variables for the sidebands. Exact analytic cancellations can turn into near cancellations numerically, requiring an extremely dense radial grid. To avoid this, we change the variable for the sidebands and solve for $\bar{\xi}^{(m\pm1)}$. It is related to the original sideband plasma displacement by: $\xi^{(m\pm1)} = \bar{\xi}^{(m\pm1)} + \frac{1\pm m}{r^{2\pm m}}\int_0^r l^{2\pm m}\Delta' \frac{d}{dl}\left(\frac{\Delta q}{q_s}\xi^{(m)} - \chi\right) dl$. We detail below the equations corresponding to each line.

The first line of (B1) is the governing equation for the main harmonic of radial plasma displacement $\xi^{(m)}$, given below for $m \neq 1$ (where shaping is now included in the Mercier term $D_I(r)$ given in (1)):

$$
\begin{aligned}
0 =& \frac{\gamma^2(1+2q_s^2)}{m^2\omega_A^2}\left\{\frac{1}{r}\frac{d}{dr}\left[r^3\frac{d\xi^{(m)}}{dr}\right]-(m^2-1)\xi^{(m)}\right\}+\frac{1}{r}\frac{d}{dr}\left[r^3\left(\frac{1}{q}-\frac{1}{q_s}\right)^2\frac{d\xi^{(m)}}{dr}\right]-(m^2-1)\left(\frac{1}{q}-\frac{1}{q_s}\right)^2\xi^{(m)}\\
&+\frac{1}{r}\frac{d}{dr}\left[\frac{r^3}{q_s}\left(\frac{1}{q}-\frac{1}{q_s}\right)\frac{d\chi}{dr}-\frac{r^3}{q_s}\chi\frac{d}{dr}\left(\frac{1}{q}\right)\right]-(m^2-1)\frac{1}{q_s}\left(\frac{1}{q}-\frac{1}{q_s}\right)\chi\\
&+\frac{s^2}{q_s^2}D_I(r)\,\xi^{(m)}\\
&+\frac{\alpha\epsilon}{q_s^2}\left(\frac{1}{q_s^2}-1\right)\Delta\xi_\Gamma-\frac{\alpha^2}{2q_s^2}\left(\xi^{(m)}+\Delta\xi_\Gamma\right)\\
&+\frac{2+m}{2(1+m)q_s^2}\left(\alpha-2\frac{\Delta q}{q_s}\left[(1+m)\epsilon+\alpha+m\Delta'\right]\right)\bar{\xi}^{(m+1)}\\
&+\frac{2-m}{2(1-m)q_s^2}\left(\alpha-2\frac{\Delta q}{q_s}\left[(1-m)\epsilon+\alpha-m\Delta'\right]\right)\bar{\xi}^{(m-1)}\\
&+\frac{r}{2(1+m)q_s^2}\left(\alpha-2\frac{\Delta q}{q_s}\left[2(1+m)\epsilon+(2+m)\alpha-(1+2m)\Delta'\right]\right)\bar{\xi}^{(m+1)\prime}-\frac{\Delta q}{q_s^3}\frac{r^2}{(1+m)}\Delta'\bar{\xi}^{(m+1)\prime\prime}\\
&+\frac{r}{2(1-m)q_s^2}\left(\alpha-2\frac{\Delta q}{q_s}\left[2(1-m)\epsilon+(2-m)\alpha-(1-2m)\Delta'\right]\right)\bar{\xi}^{(m-1)\prime}-\frac{\Delta q}{q_s^3}\frac{r^2}{(1-m)}\Delta'\bar{\xi}^{(m-1)\prime\prime}\\
&+\chi\left(-\frac{1}{q_s^2}\left[4\epsilon^2(2-\frac{1}{q_s^2})+\epsilon\alpha(3+\frac{1}{q_s^2})+\alpha^2+\Delta'(12\Delta'-7\alpha+r\alpha'-6\epsilon)\right]\right)\\
&+\xi^{(m)}\left(\frac{\Delta q}{q_s^3}\left[4\epsilon^2(2-\frac{1}{q_s^2})+\epsilon\alpha(5-\frac{2}{q_s^2})+2\alpha^2+\Delta'(12\Delta'-7\alpha+r\alpha'-6\epsilon)\right]\right)\\
&+\frac{r}{q_s^2}\alpha\Delta'\left(\frac{\Delta q}{q_s}\xi^{(m)\prime}-\chi'\right)\\
&+\Delta\xi_\Gamma\left(\frac{\Delta q}{q_s^3}\left[\epsilon\alpha\left(2-\frac{3}{q_s^2}\right)+\alpha^2\right]-\frac{q'}{q_s^3}r\Delta'\alpha\right).
\end{aligned}
\tag{B2}
$$

The $m = 1$ version reads:

$$
\begin{aligned}
0 =& \frac{\gamma^2(1+2q_s^2)}{m^2\omega_A^2}\left\{\frac{1}{r}\frac{d}{dr}\left[r^3\frac{d\xi^{(m)}}{dr}\right]\right\} + \frac{1}{r}\frac{d}{dr}\left[r^3\left(\frac{1}{q}-\frac{1}{q_s}\right)^2\frac{d\xi^{(m)}}{dr}\right] \\
&+ \frac{1}{r}\frac{d}{dr}\left[\frac{r^3}{q_s}\left(\frac{1}{q}-\frac{1}{q_s}\right)\frac{d\chi}{dr} - \frac{r^3}{q_s}\chi\frac{d}{dr}\left(\frac{1}{q}\right)\right] \\
&+ \frac{\alpha}{q_s^2}\left[\epsilon\left(\frac{1}{q_s^2}-1+\frac{\Delta q}{2q_s}\right)-\frac{\alpha}{4}\right]\left(\xi^{(m)}+\Delta\xi_\Gamma\right) + \frac{\alpha\epsilon}{q_s^2}\left[\frac{3}{4}(\kappa-1)\left(1-\frac{2}{\epsilon}\right)\right]\xi^{(m)} \\
&- \frac{r\alpha\Delta'}{2q_s^2}\left(2\chi' + \frac{\Delta q'}{q_s}\Delta\xi_\Gamma + \frac{\Delta q}{q_s}\left(\Delta\xi_\Gamma' - \xi^{(m)\prime}\right)\right) \\
&+ \frac{2+m}{2(1+m)q_s^2}\left(\alpha - 2\frac{\Delta q}{q_s}\left[(1+m)\epsilon+\alpha+m\Delta'\right]\right)\bar{\xi}^{(m+1)} \\
&+ \frac{r}{2(1+m)q_s^2}\left(\alpha - 2\frac{\Delta q}{q_s}\left[2(1+m)\epsilon+(2+m)\alpha-(1+2m)\Delta'\right]\right)\bar{\xi}^{(m+1)\prime} - \frac{\Delta q}{q_s^3}\frac{r^2}{(1+m)}\Delta'\bar{\xi}^{(m+1)\prime\prime} \\
&+ \chi\left(-\frac{1}{q_s^2}\left[4\epsilon^2\left(2-\frac{1}{q_s^2}\right)+\epsilon\alpha\left(\frac{7}{2}+\frac{1}{q_s^2}\right)+\frac{3}{2}\alpha^2+\Delta'(12\Delta'-7\alpha+r\alpha'-6\epsilon)\right]\right) \\
&+ \xi^{(m)}\left(\frac{\Delta q}{q_s^3}\left[4\epsilon^2\left(2-\frac{1}{q_s^2}\right)+\epsilon\alpha\left(5-\frac{2}{q_s^2}\right)+\frac{3}{2}\alpha^2+\Delta'\left(12\Delta'-\frac{13}{2}\alpha+\frac{1}{2}r\alpha'-6\epsilon\right)\right]\right) \\
&+ \frac{\Delta q}{q_s^3}\Delta\xi_\Gamma\left(3\epsilon\alpha\left(\frac{1}{2}-\frac{1}{q_s^2}\right)+\frac{1}{2}\Delta'(\alpha-r\alpha')\right) - \frac{\Delta q'}{2q_s^3}\Delta\xi_\Gamma r\Delta'\alpha\,.
\end{aligned}
\tag{B3}
$$

The modified sidebands also follow governing equations (corresponding to the 3rd and 5th lines of (B1)). They are given below in eq. (B4), with $q_{m\pm1} \equiv (m\pm1)/n$:

$$
\begin{aligned}
0 =& \frac{1}{r}\frac{d}{dr}\left[r^3\left(\frac{1}{q}-\frac{n}{m\pm1}\right)^2\frac{d\bar{\xi}^{r(m\pm1)}}{dr}\right] - m(m\pm2)\left(\frac{1}{q}-\frac{n}{(m\pm1)}\right)^2\bar{\xi}^{r(m\pm1)} \\
&+ \frac{\gamma^2(1+2q_{m\pm1})^2)}{(m\pm1)^2\omega_A^2}\left\{\left(\frac{1}{r}\frac{d}{dr}\left[r^3\frac{d\bar{\xi}^{r(m\pm1)}}{dr}\right] - m(m\pm2)\bar{\xi}^{r(m\pm1)}\right)\right\} \\
&+ \frac{1}{r}\frac{d}{dr}\left[r^3\frac{n}{(m\pm1)}\left(\frac{1}{q}-\frac{n}{(m\pm1)}\right)\frac{d\chi^{(m\pm1)}}{dr}\right] - \frac{1}{r}\frac{d}{dr}\left[r^3\frac{n}{(m\pm1)}\left(\frac{1}{q}\right)'\chi^{(m\pm1)}\right] \\
&- m(m\pm2)\frac{1}{q_{m\pm1}}\left(\frac{1}{q}-\frac{n}{(m\pm1)}\right)\chi^{(m\pm1)} \\
&- \frac{r^{1\pm m}}{2q_s^2(1\pm m)}\frac{d}{dr}\left(\frac{\alpha}{r^{\pm m}}\left(\xi^{(m)}+\Delta\xi_\Gamma\right)\right) \\
&+ \frac{r^{1\pm m}}{q_s^2(1\pm m)}\frac{d}{dr}\left(\left[(1\pm2m)\epsilon+(1\pm m)(\alpha-3\Delta')\right]\frac{1}{r^{\pm m}}\left(\frac{\Delta q}{q_s}\xi^{(m)}-\chi^{(m)}\right)\right) \\
&+ \frac{1}{q_s^2}(2\pm m)(\epsilon+\alpha-4\Delta')\left(\frac{\Delta q}{q_s}\xi^{(m)}-\chi^{(m)}\right) \\
&+ \frac{r^{1\pm m}}{q_s^2(1\pm m)}\frac{d}{dr}\left[\frac{\alpha}{r^{\pm m}}\left(\chi^{(m)}+\frac{\Delta q}{q_s}\Delta\xi_\Gamma\right)\right].
\end{aligned}
\tag{B4}
$$

Ohm's law enters the problem in the 2nd, 4th, and 6th lines of (B1). We write it here in eq. (B5) for the main harmonic component, and in eq. (B6) for the sidebands components.

$$\frac{\gamma}{\omega_A} r^3 \chi = \frac{r_s^2}{S_L}\left[\frac{d}{dr}\left(r^3 \frac{d}{dr}\left(\frac{m}{n}\left(\frac{1}{q} - \frac{n}{m}\right)\xi + \chi\right)\right) + r(1-m^2)\left(\frac{m}{n}\left(\frac{1}{q} - \frac{n}{m}\right)\xi + \chi\right)\right]. \tag{B5}$$

$$\frac{\gamma}{\omega_A} r^3 \chi^{(m\pm1)} = \frac{r_{m\pm1}^2}{S_L}\left[\frac{d}{dr}\left(r^3 \frac{d}{dr}\left(\frac{(m\pm1)}{n}\left\{\frac{1}{q} - \frac{n}{(m\pm1)}\right\}\overline{\xi}^{(m\pm1)} + \chi^{(m\pm1)}\right)\right) + r(1-(m\pm1)^2)\left(\frac{(m\pm1)}{n}\left\{\frac{1}{q} - \frac{n}{(m\pm1)}\right\}\overline{\xi}^{(m\pm1)} + \chi^{(m\pm1)}\right)\right]. \tag{B6}$$

Finally, the last line of (B1) corresponds to the definition of $\Delta\xi_\Gamma$:

$$\frac{\gamma^2}{\omega_A^2}\Delta\xi_\Gamma \omega_A \frac{q}{n^2} = -\omega_s^2 \Delta q\left(\chi + \frac{\Delta q}{q}\Delta\xi_\Gamma\right). \tag{B7}$$

The operators in (B1) are discretized using a finite difference method, and the generalized eigenvalue problem is solved on an adaptive grid refined around the rational surfaces of the main harmonic and the sidebands.

## Appendix C: Calculations for the energy approach

To compute the left-hand side of equation (11), we consider that it is dominant within the singular layer around the rational surface. We therefore minimize only this term with respect to the displacement, leading to:

$$\frac{\mathrm{d}}{\mathrm{d}r}\left(r^{2m+1}F(r)^2\frac{\mathrm{d}}{\mathrm{d}r}\left(r^{1-m}\xi\right)\right) = 0\,, \tag{C1}$$

where $F(r)^2 = \hat{\gamma}^2 + (\frac{1}{q} - \frac{1}{q_s})^2$. We solve (C1) within the layer to obtain $\frac{\mathrm{d}}{\mathrm{d}r}\left(r^{1-m}\xi\right)|_{r=r^*}$. We introduce a dimensionless layer variable $x \equiv \frac{r-r^*}{r^*}$, and the equation (C1) becomes

$$\frac{\mathrm{d}}{\mathrm{d}x}\left(\left(\hat{\gamma}^2 + (\frac{1}{q} - \frac{1}{q_s})^2\right)\frac{\mathrm{d}}{\mathrm{d}x}\xi\right) = 0\,,$$

which can be rewritten as:

$$\begin{aligned}
\left[\hat{\gamma}^2 + \left(\frac{1}{q} - \frac{1}{q_s}\right)^2\right]\frac{\mathrm{d}}{\mathrm{d}x}\xi &= \delta W\,\xi_0 \\
\frac{\mathrm{d}}{\mathrm{d}x}\xi &= \frac{\delta W \xi_0}{(\hat{\gamma}^2 + \left(\frac{1}{q} - \frac{1}{q_s}\right)^2)} \\
0 - \xi_0 &= \int_{-\infty}^{+\infty}\frac{\delta W \xi_0}{(\hat{\gamma}^2 + \left(\frac{1}{q} - \frac{1}{q_s}\right)^2)}\mathrm{d}x \\
\delta W &= -\left[\int_{-\infty}^{+\infty}\frac{1}{(\hat{\gamma}^2 + \left(\frac{1}{q} - \frac{1}{q_s}\right)^2)}\mathrm{d}x\right]^{-1}\,,
\end{aligned} \tag{C2}$$

where we extend the integral bounds to $\pm\infty$ since the integrand vanishes away from the singular layer. We have $\xi(x = +\infty) = 0$, and we define $\xi(x = -\infty) = \lim_{\delta_r \to 0^+}\xi(r_\star - \delta_r) \equiv \xi_0$, the value just before the end of the low shear zone. To evaluate the integral in (C2), we expand $q$ around $r_\star$ as: $q = q_*(1 + s_* x + (q_*'' r_\star^2/2q_\star)x^2)$, and we obtain:

$$\left(\frac{1}{q} - \frac{1}{q_s}\right)^2 = \frac{1}{q_*^2 q_s^2}\left(\Delta q^2 + 2\Delta q f_1 x + x^2(f_1^2 + 2\Delta q f_2)\right) + O(x^5)\,. \tag{C3}$$

Keeping terms up to $O(x^2)$ in the primitive on the right-hand side of eq. (C2) (to ensure convergence), and rearranging to solve for , we obtain:

$$\delta W = -\frac{\sqrt{\hat{\gamma}^2 + \frac{\Delta q^2}{q_*^2 q_s^2}}\sqrt{\frac{s_*^2}{q_*^2} + \frac{\Delta q q_*'' r_*^2}{q_s^2 q_*^2} - \frac{2\Delta q s_*^2}{q_s^2 q_*}}}{\pi} .$$

As a consistency check for the general validity of our ansatz for $\xi$ (given in eq. (14)), we verify a posteriori that the first line of eq. (C2) holds. This is illustrated in Figure 22, for the most unstable root of the $m = n = 1$ case presented in Fig. 4 ($\gamma/\omega_A = 0.0195$, and $r_\star = 0.3$). The quantity of interest is not exactly zero throughout the region $r < r^*$, but remains small enough to give us confidence in the choice of ansatz.

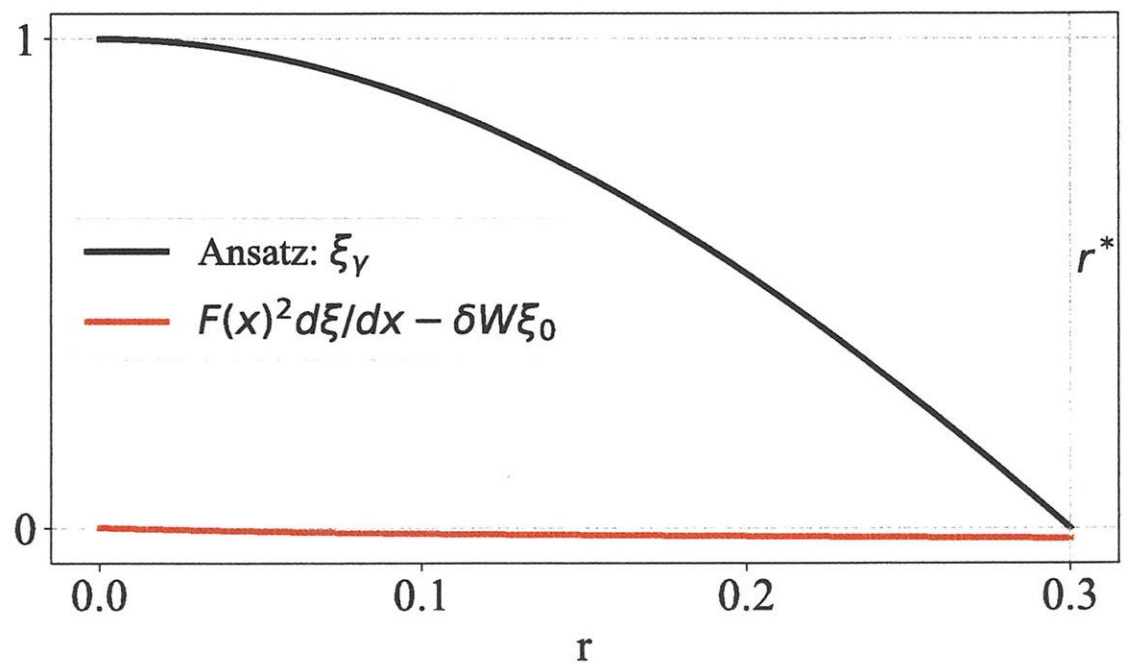

Figure 22: Checking the validity of the ansatz $\xi_\gamma$ from eq. (14): the quantity $\left[\hat{\gamma}^2 + \left(\frac{1}{q} - \frac{1}{q_s}\right)^2\right]\frac{\mathrm{d}}{\mathrm{d}x}\xi - \delta W\,\xi_0$ should be equal to zero. $m = n = 1$ case of Figure 4.

## Appendix D: Profiles used

In this work, we set $R_0 = 3$ (major radius), $a = 1$ (value of the minor radius $r$ at the edge), and $B_0 = 3T$ (on-axis value of the magnetic field, required for the resistive results). Unless stated otherwise, a parabolic $\beta$ profile is used throughout:

$$\beta = \beta_0(1 - r^2) ,$$

with $\beta_0 \equiv \alpha(r_s)/(2q(r_s)^2 R_0 r_s)$. For figures 18 and 19 a peaked pressure profile is chosen :

$$\beta_0 = -(\beta_0/2)\tanh\left(10\beta_1(r - 0.5/\beta_1)\right) + \beta_0/2 , \tag{D1}$$

with $\beta_0 = 0.1$ and varying $\beta_1$ to obtain desired value of $\alpha$ on the rational. For these figures 18 and 19, an ultra flat $q$ profile is chosen:

$$q = (q_s - \Delta q)(1 + c\,r^{2d})^{1/d} , \tag{D2}$$

with $q_s = 9/10$, $r_s = 0.3$ and $q_s = 5$. $q_0 = q(r = 0)$, $c$ and $d$ vary to adjust magnetic shear.